\documentclass[useAMS,usenatbib,usegraphicx]{mn2e}
\usepackage{psfig,hyperref,amssymb,subfigure}
\newcommand\beq{\begin{equation}}
\newcommand\eeq{\end{equation}}

\def\msun{\,{\rm M_\odot}}

\def\gsim{ \lower .75ex \hbox{$\sim$} \llap{\raise .27ex \hbox{$>$}} }
\def\lsim{ \lower .75ex\hbox{$\sim$} \llap{\raise .27ex \hbox{$<$}} }

\title[Multimessenger astronomy with pulsar timing]
{Multimessenger astronomy with pulsar timing and X-ray observations of massive black hole binaries}

\author[A. Sesana et al.]{A. Sesana$^{1}$, C. Roedig$^{1}$, M.T. Reynolds$^{2}$ and M. Dotti$^{3,4}$ \\
$^{1}$ Max-Planck-Institut f{\"u}r Gravitationsphysik, Albert Einstein
Institut, Am M{\"u}hlenberg 1, 14476 Golm, Germany \\
$^{2}$ Department of Astronomy, University of Michigan, 500 Church Street, Ann Arbor,
MI 48109, USA\\
$^3$ Max-Planck-Institut f{\"u}r Astrophysik, Karl-Schwarzschild-Stra{\ss} e 1, 
D-85748 Garching b. M\"unchen, Germany\\
$^4$ Universit\`a di Milano Bicocca, Dipartimento di Fisica
G. Occhialini, Piazza della Scienza 3, I-20126, Milano, Italy\\
}

\begin{document}
\date{Received ---}

\maketitle

\begin{abstract}
In the decade of the dawn of gravitational wave astronomy, the concept of multimessenger astronomy, combining gravitational wave signals to conventional electromagnetic observation, has attracted the attention of the astrophysical community. So far, most of the effort has been focused on ground and space based laser interferometer sources, with little attention devoted to the ongoing and upcoming pulsar timing arrays (PTAs). We argue in this paper that PTA sources, being very massive ($>10^8\msun$), cosmologically nearby ($z<1$) black hole binaries (MBHBs), are particularly appealing multimessenger carriers. According to current models for massive black hole formation and evolution, the planned Square Kilometer Array (SKA) will observe thousands of such massive systems, being able to individually resolve and locate in the sky several of them (maybe up to a hundred). MBHBs form in galaxy mergers, which are usually accompanied by strong inflows of gas in the center of the merger remnant. By employing a standard model for the evolution of MBHBs in circumbinary discs, with the aid of dedicated numerical simulations, we characterize the gas-binary interplay, identifying possible electromagnetic signatures of the PTA sources. We concentrate our investigation on two particularly promising scenarios in the high energy domain, namely, the detection of X-ray periodic variability and of double broad K$\alpha$ iron lines. Up to several hundreds of periodic X-ray sources with a flux $>10^{-13}$erg s$^{-1}$cm$^{-2}$ will be in the reach of upcoming X-ray observatories; in the most optimistic case, few of them may be already being observed by the MAXI detector placed on the International Space Station. Double relativistic K$\alpha$ lines may be observable in a handful of low redshift ($z<0.3$) sources by proposed deep X-ray probes, such as \textit{Athena}. The exact figures depend on the details of the adopted MBHB population and on the properties of the circumbinary discs, but the existence of a sizable population of sources suitable to multimessenger astronomy is a robust prediction of our investigation.

\end{abstract}


\section{Introduction}
Within this decade the detection of gravitational waves (GWs) may be a reality,
opening a completely new window on the Universe.  While signals coming from compact
stars and binaries fall in the observational domain of operating and planned ground
based interferometers (such as LIGO, VIRGO, and the proposed Einstein Telescope),
massive black hole (MBH) binaries (MBHBs) are expected to be among the primary actors
on the upcoming low frequency stage, where the $10^{-4}-10^{-1}$Hz window is going to
be probed by gravitational wave interferometry in space (see, e.g., the proposed Laser
Interferometer Space Antenna (LISA)). Moving to even lower frequencies, long term
monitoring of an array of millisecond pulsars (forming a so called pulsar timing
array, PTA) may unveil the characteristic fingerprint left by GWs in the time of
arrival of the radio pulses \citep[e.g.,][]{sazhin78,hellings83}.  The Parkes Pulsar
Timing Array \citep[PPTA,~][]{manchester2008}, the European Pulsar Timing Array
\citep[EPTA,~][]{janssen2008} and the North American Nanohertz Observatory for
Gravitational Waves \citep[NANOGrav,~][]{jenet2009}, joining together in the
International Pulsar Timing Array \citep[IPTA,~][]{hobbs2010}, are already collecting
data and improving their sensitivity in the frequency range of $\sim10^{-9}-10^{-6}$
Hz, and in the next decade the planned Square Kilometer Array
\citep[SKA,~][]{lazio2009} will provide a major leap in sensitivity.  Even though GW
observations alone will be an outstanding breakthrough in science, the astrophysical
payouts will be greatly amplified by the coincident identification of electromagnetic
counterparts. The identification of the host galaxy of a MBH binary merger will (i)
improve our understanding of the nature of the galaxy hosting coalescing MBHBs
(e.g. galaxy type, colours, morphology, etc.); (ii) help to reconstruct the dynamics
of the merging galaxies and of their MBHs; and (iii) offer the possibility of studying
accretion phenomena onto systems of known mass and spin \citep[which, also in PTA
  observations, can be measured from the GW signal,
  e.g.][]{vecchio2004,SesanaVecchio10,corbin2010}.

Electromagnetic counterparts to GW events have attracted a lot of attention in the
last few years in the context of LISA observations of coalescing MBHBs \citep[see][and
  references therein]{schnittman2011}. The final coalescence being a violent event,
many possible counterparts related to shocks flares and transients in the surrounding
ambient gas have been proposed.
However, the bulk of the LISA sources, are likely to cluster at $3<z<8$
\citep{sesana2007}.  Given the modest sky localization accuracy of the detector
\citep[generally $>1$deg$^2$,][]{lang2008}, the number of candidate hosts in the
putative GW errorbox is likely to be very high (order of millions).  Moreover, our
incomplete understanding of the observational signatures of such events, together with
their intrinsic weakness \citep[most LISA sources are expected to be MBHBs with masses
  $\lesssim10^6\msun$][]{sesana2007} will make host identification extremely
challenging.

On the other hand, very little attention has been devoted to possible counterparts of
PTA sources. An obvious advantage with respect to LISA sources is that those are
expected to be very massive ($M>10^8\msun$), cosmologically nearby ($z<1$) systems
\citep{SVV09}. Any characteristic electromagnetic signature would be therefore within
the sensitivity range of current and future observation capabilities.  PTA sources are
inspiralling MBHBs still far from coalescence (emitting in the $\sim10^{-9}-10^{-6}$
Hz frequency range), that can be considered, for any practical purpose, stationary
during typical observation timescales of decades \citep{SesanaVecchio10}. Moreover,
fairly high eccentricity should be the norm, as shown by MBHB hardening studies in
both stellar \citep{Sesana2010,SGD11,Khan11,Preto11} and gaseous
\citep{Armitage:2005,Jorge09,roedig11} environments. If the binary is surrounded by a
circumbinary gaseous disc, eccentricity triggers periodic inflows
\citep{arty96,roedig11}, opening a wide range of possibility for distinctive binary
fingerprints that can be observationally identified.

In this paper we study the prospects of conducting multimessenger astronomy with PTA
observations of MBHBs. Our aim is to quantify the number of sources possibly
detectable both in the GW and in the electromagnetic realm, identifying their
signatures, to forecast future identification.  Starting from the Millennium Run
\citep{springel}
database{\footnote{http://www.mpa-garching.mpg.de/galform/virgo/millennium/}}, we
construct detailed MBHB populations satisfying all the currently available constraints
in terms of local MBH mass function \citep{marconi2004}, MBH-host relations
\citep{gultekin09}, and MBHB merger rates as inferred from close galaxy pair counts
\cite[e.g.][]{bell2006}.  We assume that after a galaxy merger, cold gas is funneled
in the center of the remnant, forming a circumnuclear disc
\citep{mihos96,Mayer07}. The new-formed sub-parsec MBHB excavates a hollow region in
the center of the disc (usually referred to as 'gap'), and the dynamics of the system
is governed by the mutual MBHB-disc torques. We compute the detailed population of GW
emitting sources at each frequency by adopting a simple analytical model for their
evolution under the dynamical effect of gaseous drag and GW emission \cite[see][for
  details]{KS11}. After identifying the prototypical MBHB observable with PTAs, we
simulate its dynamics numerically using a modified version of the smooth particle
hydrodynamics (SPH) code GADGET-2 \citep{Springel05,Jorge09}, to track the fate of the
material leaking through the gap and captured by the MBHs. The outcome of the
simulation are then used to model analytically the accretion onto the two MBHs and the
emitted radiation.

Several distinctive signatures of inspiralling MBHBs have been presented in the
literature.  Proposed scenarios range from the radio and the optical, up to the
X-rays, including variability of the optical continuum \citep[e.g.,][]{Haiman2009b},
spectral shifts in the broad line emission \citep{begelman80,Eracleous11,Tsalmantza11}, 
peculiar flux ratios between optical/UV broad emission lines \citep{montuori11}, 
precessing or x-shaped jets \citep{liufukun2004,lobanov2005}, periodic outbursts
\citep{valtonen1988}, astrometric measurement of the source motion
\citep{sudou2003}. Yet, the only unambiguous candidate is a double compact radio core
at a 7pc projected separation \cite{Rodriguez2006}.  We identify here a number of
possible characteristic signatures and we concentrate our investigation on two
particularly promising scenarios in the high energy domain, namely, the detection of
X-ray periodic variability and of double broad K$\alpha$ iron lines. For both
scenarios, we quantify the population of potentially observable sources, and the joint
detection prospects with future X-ray observatories and PTAs.

The paper is organized as follows. In Section 2 and 3 we model and describe in detail
the MBHB population relevant to PTA observations; whereas in Section 4 we investigate
the dynamics of the typical source by combining high resolution SPH simulations to an
analytical model for the accreted matter. In Section 5 we model the emitted spectrum,
identifying possible characteristic signatures, and Section 6 and 7 are devoted to
periodic variability and double broad K$\alpha$ lines, respectively.  We discuss
strategy for joint PTA and X-ray detection and draw our conclusions in Section 8.

\section{The pulsar timing massive black hole binary population}
Our main goal is to quantify and characterize the population of MBHBs that is
accessible to observation both in GWs through pulsar timing and in the electromagnetic
realm via distinctive signatures. Formally, what we need to estimate is the population
of MBHBs observable today in the universe as a function of MBH masses, redshift and
orbital frequency (or, alternatively, semimajor axis): $d^4N/(dM_1dM_2dzd{\rm
  ln}f_{r,k})$. Here $M_1>M_2$ are the masses of the two binary components, $z$ is the
source redshift and $f_{r,k}$ is the rest-frame Keplerian frequency of the
system. This can be formally written as the product of (i) $d^4N/(dM_1dM_2dzdt_r)$,
times (ii) $dt_r/d{\rm ln}f_{r,k}$. Item (i) is nothing else but the cosmological
coalescence rate of MBHBs, and it depends on the general clustering history of
structures in the Universe; item (ii) describes the time evolution of the binary
frequency, and is determined by the detailed physical processes driving the binary
dynamics prior to coalescence \footnote{For a given coalescence rate, $dt_r/d{\rm ln}f_{r,k}$
quantifies the number of binaries as a function of orbital frequency that has to be 
present at any time in the sky to sustain that particular rate. This item can be 
estimated taking into account that we are interested in subparsec MBHB, for which 
we can employ simple evolutionary models driven by disc-binary mutual torques and 
GW emission, as we will see in Section 2.2.}. 
Once such a population has been determined, the
requirement of observability through PTA, selects a subset of systems which we will
refer to as the PTA--MBHB population. In the following we will treat items (i)-(ii)
separately (Sections \ref{mbhbpop} and \ref{dyn}), and we will describe the
characteristic timing residuals induced in a PTA, defining the PTA--MBHB population
(Section \ref{ptaobs}).

\subsection{Cosmological coalescence rate}
\label{mbhbpop}
To construct the MBHB cosmological coalescence rate, we follow the procedure described
in Section 2 of \cite{SVV09}, the reader is deferred to that paper for full
details. In few words, we extract catalogs of merging galaxies from the
semi-analytical model of \cite{Bertone07} applied to the Millennium run
\citep{springel}.  We then associate a central MBH to each merging galaxy in our
catalogue. We adopt the recent fit to the $M-\sigma$ relation given by
\cite{gultekin09}, and we consider accretion to be efficient onto both MBHs {\it
  before} the final coalescence of the binary.

Assigning a MBH to each galaxy, we obtain a catalogue of all the mergers occurring in
the $(500/h {\rm Mpc})^3$ comoving volume of the simulation, labeled by MBH masses and
redshift. From this, we numerically generate the cosmic merger rate, $d^4
N/(dM_1dM_2dzdt_r)$, by weighting each event with the observable comoving volume shell
at every redshift.

\subsection{MBH binaries in circumbinary discs}
\label{dyn}
We assume that MBHBs evolve in geometrically thin circumbinary accretion discs
\citep{Haiman2009}. Even though such an assumption is enforced by the requirement of a
distinctive electromagnetic signature, we find that most of the merging galaxy pairs
extracted by the Millennium run involve at least one massive, gas rich, spiral
galaxy. Significant inflows of cold gas are therefore expected in the remnant nuclei
\citep{mihos96}, feeding the circumbinary disc.

Before entering the dynamics in detail, we define all the quantities used for
describing the MBHB-disc system. The MBHB has total mass $M=M_1+M_2$ ($M_1>M_2$), mass
ratio $q=M_1/M_2$, symmetric binary mass ratio $q_{\rm s}=4q/(1+q^2)$, semimajor axis
$a$, rest-frame orbital Keplerian frequency $f_{k,r}$, and eccentricity $e$. The thin
disc is described in terms of the $\alpha$ viscosity parameter, the accretion rate
$\dot{M}$ and the accreted matter/radiation conversion efficiency
$\epsilon$. Following \cite{roedig11}, we define $\delta$ to be the relative size of
the gap to the binary semimajor axis, and we take a fiducial value $\delta=2$.
Throughout the paper, expressions are given in units of $M_8=M/10^8\msun$,
$\alpha_{0.3}=\alpha/0.3$, $\epsilon_{0.1}=\epsilon/0.1$ and
$\dot{m}_{0.3}=\dot{m}/0.3$, where $\dot{m}=\dot{M}/\dot{M}_{\rm Edd}$ is the
accretion rate normalized to the Eddington rate. Subscript '3' refers to lengths given
in units of $10^3R_S$ where $R_s=2GM/c^2$ is the Schwarzschild radius associated to
the total mass of the binary.

The picture we adopt for the MBHB evolution is the following. At separations relevant
to our work ($a<0.03$ pc), the binary has excavated a cavity \citep[gap, see,
  e.g.][]{Artymowicz1994} in the circumbinary accretion disc. Both viscous torques
exerted by the disc \citep{GoldreichTremaine80} and GW emission \citep{Peters:1963ux}
dissipate the binary binding energy causing its orbital decay. The secondary MBH is,
in general, more massive than the local disc and the problem is analogous to
secondary-dominated Type-II migration in planetary dynamics. A self-consistent
solution to this problem was provided by \cite{syer95}, the interested reader is
referred to their original paper and to \cite{Haiman2009} and \cite{KS11} for a
detailed discussion in the context of MBHB evolution. Such a solution applies only to
discs with decreasing surface density as a function of the radius. Standard
$\alpha$-discs \citep{ss73}, with viscosity proportional to the total pressure, meet
such a requirement only in the gas pressure dominated zone, but not in the inner
radiation pressure dominated zone. $\beta$-discs, with viscosity proportional to the
gas pressure only, fulfill this condition at all radii. We therefore assume
$\beta$-discs as our fiducial disc model for describing the binary-disc migration. All
the relevant features of $\beta$-discs are reviewed by \cite{Haiman2009}, here we
merely introduce the quantities relevant to the practical computation of the PTA-MBHB
population.  The relevant timescales in the system are:\\ (i) the viscous timescale
\begin{equation}
t_{\nu}=6.09\times10^5 \,{\rm yr}\, \alpha_{0.3}^{-4/5}\left(\frac{\dot{m}_{0.3}}{\epsilon_{0.1}}\right)^{-2/5}M_8^{6/5}\delta^{7/5} a_3^{7/5};
\label{tnu}
\end{equation}
(ii) the migration timescale
\begin{equation}
t_m=2.09\times10^6 \,{\rm yr}\, \alpha_{0.3}^{-1/2}\left(\frac{\dot{m}_{0.3}}{\epsilon_{0.1}}\right)^{-5/8}M_8^{3/4} q_{\rm s}^{3/8}\delta^{7/8} a_3^{7/8};
\label{tm}
\end{equation}
(iii) the GW shrinking timescale
\begin{equation}
t_{\rm GW}=7.84\times10^7 \,{\rm yr}\, M_8 q_{\rm s}^{-1} a_3^{4} F(e)^{-1},
\label{tgw}
\end{equation}
where
\begin{equation}
F(e)=(1-e^2)^{-7/2}\left(1+\frac{73}{24}e^2 +\frac{37}{96}e^4 \right).
\end{equation}
These timescales define two characteristic binary separations. Equating
$t_m$ to $t_{\rm gw}$ we get the decoupling separation
\begin{equation}
a_3^{\rm dec}=0.31\,\alpha_{0.3}^{-4/25}\left(\frac{\dot{m}_{0.3}}{\epsilon_{0.1}}\right)^{-1/5}M_8^{-2/25} q_{\rm s}^{11/25}\delta^{7/25} F(e)^{8/25},
\label{adec}
\end{equation}
which is the separation below which the GW driven migration is faster than the gas driven migration, and
the binary evolution is dictated by GW emission only. This is the relevant separation that has to
be considered when computing the frequency distribution of inspiralling MBHBs. Equating $t_\nu$ to
$t_{\rm gw}$ we get the 'disc freezing' (or detachment) separation
\begin{equation}
a_3^{\rm fr}=0.22\,\alpha_{0.3}^{-4/13}\left(\frac{\dot{m}_{0.3}}{\epsilon_{0.1}}\right)^{-2/13}M_8^{1/13} q_{\rm s}^{4/13}\delta^{7/13} F(e)^{5/13},
\label{afr}
\end{equation}
below which the binary shrinking timescale is faster than the viscous inward diffusion of the 
inner edge of the disc. After this point, the disc can no longer follow the binary, and it is basically 
'frozen' during the quick subsequent inspiral leading to the MBHB coalescence. This latter characteristic 
radius discriminates between
binaries attached to their discs and binaries detached from their discs, and it has to be considered
in computing the population of 'observable systems'. 
In general $a^{\rm fr}<a^{\rm dec}$, defining three different
phases relevant to our study:
\begin{itemize}
\item phaseI, $a>a^{\rm dec}$. The binary is coupled to the disc which regulates its dynamical evolution;
\item phaseII, $a^{\rm fr}<a<a^{\rm dec}$. The binary is dynamically decoupled from the disc and it is driven
by GW emission. However, the viscous time is short enough that the disc can follow the binary in its
inspiral;
\item phaseIII, $a<a^{\rm fr}$. The binary leaves the disc behind and quickly coalesces.
\end{itemize}

During phaseI, the binary maintains a constant limiting eccentricity given by
\citep{roedig11}
\begin{equation}
e_0\approx 0.66\sqrt{{\rm ln}\delta-0.65}+0.19
\label{edec}
\end{equation}
while, after decoupling, in phaseII and III, the eccentricity can be numerically computed by
solving the implicit equation
\begin{equation}
\frac{f_{\rm k,r}}{f_{\rm dec}} = \left\{ \frac{1-e_{\rm dec}^2}{1-e^2} \left(\frac{e}{e_{\rm dec}}\right)^{\frac{12}{19}} \left[\frac{1+\frac{121}{304}e^2}{1+\frac{121}{304}e_{\rm dec}^2}\right]^{\frac{870}{2299}} \right\}^{-3/2},\label{eq:f_e_relation}
\end{equation}
obtained in the quadrupole approximation by coupling the GW orbital decay rate to the eccentricity decay
rate \citep{Peters:1963ux}.

To compute the frequency distribution of inspiralling MBHBs,
following \cite{KS11}, we express the frequency evolution of the binary in terms
of 'residence time'
\begin{equation}
\left|\frac{dt_r}{d{\rm ln}f_{r,k}}\right|
= \left|\frac{dt_r}{d{\rm ln} a}\frac{d{\rm ln}a}{d{\rm ln}f_{r,k}}\right|
= \frac{2}{3}t_{\rm res}.
\label{e:dtdlnf}
\end{equation}
In phaseI, when the binary is driven by the disc, $t_{\rm res}=t_m$; while in phaseII
and phaseIII, GW emission takes over and $t_{\rm res}=t_{\rm gw}$.
Putting the pieces together, for any given MBHB-disc parameters, we
compute  $a^{\rm dec}$, assuming $e_0$ given by equation (\ref{edec});
 ${dt_r}/{d{\rm ln}f_{r,k}}$ is then obtained from equation
(\ref{e:dtdlnf}) by plugging-in equations (\ref{tm}) and (\ref{tgw}), where $e(f_{k,r})$
after decoupling is numerically obtained at any frequency from equation (\ref{eq:f_e_relation}).

\subsection{Pulsar timing observability}
\label{ptaobs}
We are interested here in MBHBs potentially observable with ongoing and forthcoming PTAs.
At the leading order, an eccentric MBHB radiates GWs in the whole spectrum of harmonics
$f_{r,n}=nf_{r,k}\,\,\,(n=1, 2, ...)$. The sky-polarization averaged amplitude
observed at a frequency $f_n=f_{r,n}/(1+z)$ at comoving distance $d$ is given by
\citep{Finn:2000sy}
\begin{equation}
h_n(f_n)=2\sqrt{\frac{32}{5}}\frac{{\cal M}^{5/3}}{nd}(2\pi f_{r,k})^{2/3}\sqrt{g(n,e)}
\label{hrms}
\end{equation}
where ${\cal M}=M_1^{3/5}M_2^{3/5}/(M_1+M_2)^{1/5}$ is the chirp mass of the system and
$g(n,e)$ is a combination of Bessel functions that quantifies the relative power
radiated in each single harmonic \citep[see][for a detailed description]{AmaroSesana10}.
If we observe for a time $T_{\rm obs}$, the relevant detectable amplitude of each harmonic
is approximately
\begin{equation}
h_{\rm obs,n}=h_n\sqrt{T_{\rm obs} f_{n}},
\label{hobs}
\end{equation}
where $\sqrt{T_{\rm obs} f_{n}}$ is simply the number of cycles completed by the
$n$-th harmonic in the observation time. 
The timing residuals
are defined as integrals of the GW during the observation time. As
detailed in \cite{SVV09}, each harmonic induces an average timing residual of
the order:
\begin{equation}
\delta t_\mathrm{gw}(f_{n}) = \sqrt{\frac{8}{15}}\frac{h_{\rm obs,n}}{2\pi f_{n}},
\label{e:deltatgws},
\end{equation}
where $\sqrt{8/15}$ is the average 'antenna beam pattern', i.e. the average of
the signal performed over all possible source-pulsar relative orientations.
The total residual can then be assumed to be of the order:
\begin{equation}
\delta t_\mathrm{gw} = \left(\sum_{n=0}^\infty\delta t_\mathrm{gw}^2(f_{n})\right)^{1/2}.
\label{e:deltatgw2}
\end{equation}

In observations with PTAs, radio-pulsars are monitored weekly for total periods of
several years \citep{hobbs2011}. Assuming a repeated observation in uniform $\Delta t$ time intervals for a
total time $T_{\rm obs}$, the maximum and minimum resolvable frequencies are
$f_{\rm max}=1/(2\Delta t) \approx 10^{-6}$Hz, corresponding to
the Nyquist frequency, and $f_{\rm min}= 1/T_{\rm obs}\approx 3\times10^{-9}$Hz for a 10 yr
observation time. We are therefore interested only in MBHBs producing a significant
residual in this frequency window.

Coupling the cosmological coalescence rate to the detailed binary evolution 
(see Sections 2.1 and 2.2) we
numerically obtain a distribution $d^4N/(dM_1dM_2dzd{\rm ln}f_{r,k})$ so that
$\int {d^4N}/({dM_1dM_2dzd{\rm ln}f_{r,k}})$ in the appropriate mass, redshift
and frequency ranges, give the 'average' number of sources observable in the
Universe taking an ideal snapshot of the whole sky at any time. To quantify the
population of relevant PTA-MBHBs, we generate Montecarlo samples of MBHBs
according to such distribution assuming $f_{r,k}>10^{-10}$Hz, $M_1 \& M_2 >10^7\msun$
and $z<3$. For each MBHB we then compute the average timing residual in the
observed frequency range $[f_{\rm min}, f_{\rm max}]$ according to equation
(\ref{e:deltatgw2}), and we keep only those binaries producing a residual
larger than 0.1ns.

\section{Description of the PTA-MBHB population}

\begin{figure*}
\includegraphics[width=\linewidth]{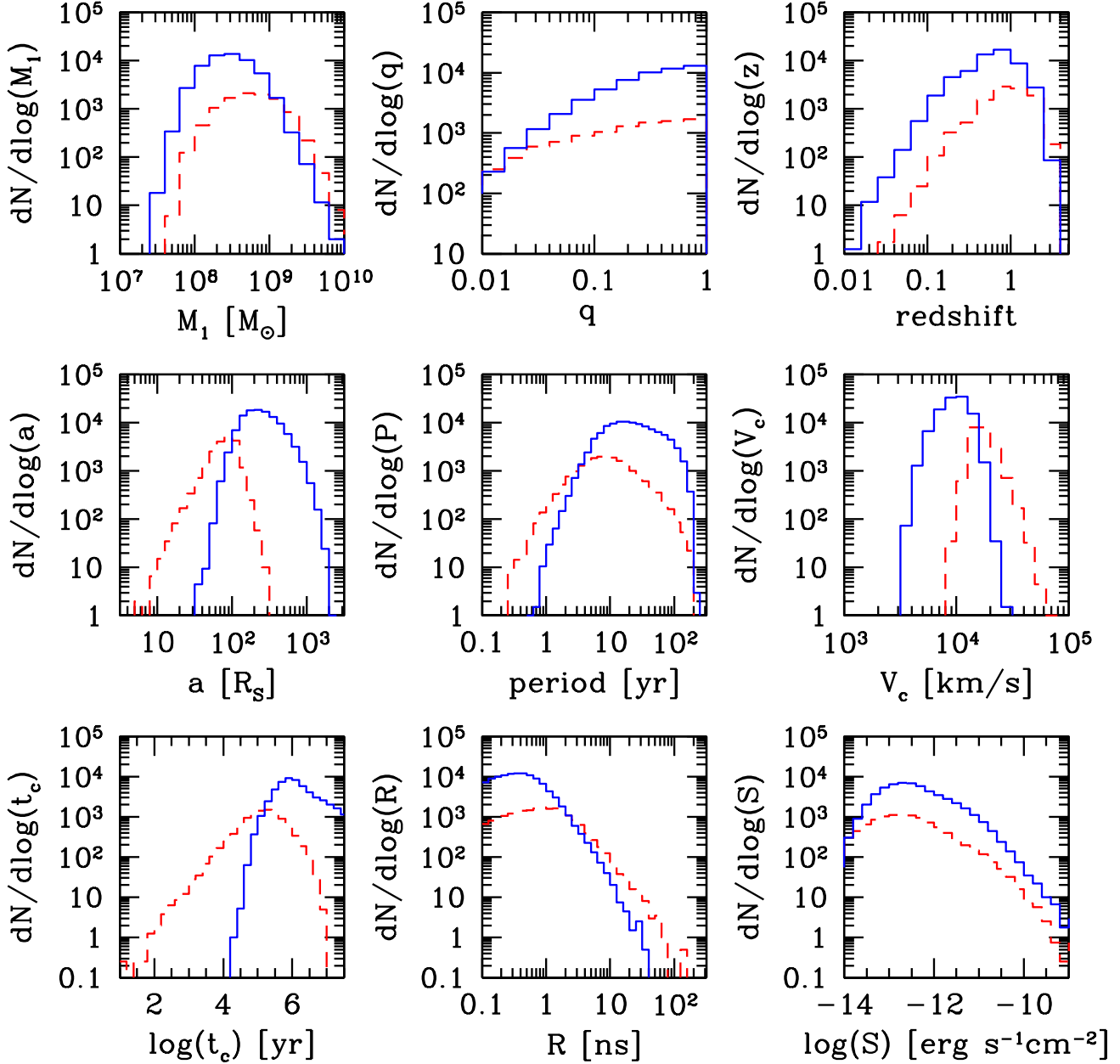}
\caption{General features of the typical MBHB population contributing at a level of
0.1ns or more to the PTA signal in the $3\times10^{-9}-10^{-6}$Hz frequency window. In each
panel the blue-solid histogram refers to sources at $a>a^{\rm fr}$, the read-dashed histogram
refers to sources at $a<a^{\rm fr}$(see text for details). In the top row, we plot
the cosmological evolution--related MBHB properties:
the primary mass $M_1$, mass ratio $q$ and redshift distributions, from the left to the right.
In the central row we plot MBHB properties describing the dynamics of the system:
the binary semimajor axis in units of $R_S$, the binary period and the circular velocity
$V_c$ distributions, from the left to the right. In the bottom row we plot the
coalescence time distribution (left), the distribution of the induced timing residuals $R$
(center) and the X-flux (0.5-10 keV) assuming a bolometric correction of $3\%$. Distribution
refers to the average over 100 Montecarlo realisations of our default model:
$\alpha=0.3, $ $\dot{m}=0.3$, $\epsilon=0.1$, and $e_{\rm dec}$ given by equation (\ref{edec}).}
\label{pop}
\end{figure*}

In our default model, we assume that MBHBs evolve according to the scheme described
in Section \ref{dyn}. We consider $\beta$-discs,
with viscosity proportional to gas pressure only and a viscosity parameter
\citep{ss73} $\alpha=0.3$ \citep[see][for a detailed discussion]{King07},
$\dot{m}=0.3$ \citep[see, e.g., ][]{Kollmeier2006,Labita09} and efficiency $\epsilon=0.1$
(appropriate for not-to-mildly rotating black holes). All the relevant features
of the resulting PTA--MBHB population, averaged over 100
Montecarlo realisations, are shown in figure \ref{pop}. Typical PTA sources are
very massive ($M>10^8$), cosmologically nearby ($z<2$) binaries. Mass
ratios are in the range 0.1-1, with a long tail extending to $10^{-3}$. Given the significant
eccentricities, orbital periods contributing to the PTA signals extend up
to $>100$ years. For the range of masses involved, this corresponds to a broad
distribution in orbital semimajor axis in the range $30-10^3R_S$ and
circular velocities peaked around $10^4$km s$^{-1}$ (if the systems were in circular orbits).
Typical coalescence times ($t_c$, defined as the integral of equation (\ref{e:dtdlnf}) from
the restframe binary frequency to coalescence) are between $10^4-10^6$ years
(bottom left panel); PTA--MBHBs are therefore caught in the final few hundred
thousand years of their life. The bottom--central panel shows the distribution
of induced residuals. Order of $\gsim 1000$ sources contribute to a level of 1ns or
more, which can realistically be considered the ultimate goal of future PTA efforts.
Their superposition will result in a confusion-noise-type foreground, similar to the
WD-WD binary signal expected for LISA \citep[see, e.g.,][]{Nelemans01}.
As in this latter case, few (maybe most) of the sources contributing to the signal may be 
individually resolvable. Preliminary, purely frequency based, crude estimations
\citep{SVV09} forecast resolvability of at least $\sim5-10$ sources, but many additional
sources will likely be resolved using the spatial information enclosed in the array
\citep[up to $2N/7$ per frequency bin, for an array of $N$ pulsars according to][the
prescription we will use below]{Boyle10}.
The bottom--right panel shows a rough estimation of the mean
X--ray flux on Earth assuming that the accretion rate of the binary (fueled by streams
flowing through the gap, see next section) is the same as the one assumed for modeling
the outer circumbinary disc (i.e., $\dot{m}=0.3$ in this case) and a bolometric
correction of $3\%$ \citep{Lusso10}. Given their high mass and low redshifts,
PTA--MBHBs are quite bright, with fluxes generally above $10^{13}$erg s$^{-1}$ cm$^{-2}$.
In all panels, blue-solid histograms refer to sources at $a>a^{\rm fr}$.
In such systems, the gas can follow the binary during its inspiral, and strongly
periodic inflows of gas feeding the two black holes are a robust prediction
of SPH simulations \citep[][see next Section]{roedig11}. Thus, we can safely
count them in the population of PTA-MBHBs suitable for
multimessenger astronomy observations. The dashed-red histograms, instead, track
binaries  at $a<a^{\rm fr}$. For such systems, torques exerted by the binary
quadrupole moment on the inner edge of the disc become smaller and smaller as
the binary shrinks, losing effectiveness in driving the periodic streaming activity.
Nonetheless, \cite{tanaka11} showed that also in this detached phase (which we 
labelled phaseIII)
gas can efficiently leak through the gap, feeding the central binary;
moreover, residual activity related to the consumption of the fossil gas left 
around the two MBHs can still be present \citep[see, e.g.,][]{chang2010}.
Therefore, detached systems should also be interesting targets for electromagnetic
counterpart identification; however, we do not consider them in the following 
discussion, and we refer the reader to \cite{tanaka11} for a comprehensive analysis of 
the subject.

\subsection{Parameter dependence and caveats}
The population shown in figure \ref{pop} relies on a selected MBH-host relation
\citep{gultekin09} and on a specific disc model (the $\beta$-disc model) with
fixed parameters. Even though the qualitative features of the predicted
population are robust against our particular choices, the effective number of sources
and the relative population of coupled/decoupled systems are not. Both lowering
$\alpha$ and $\dot{m}$ result in longer migration timescales in the disc.
This, on one hand, increases (up to a factor of a few) the number of PTA--MBHBs.
However, in this case, GW emission takes over at slightly larger separations
(see the weak dependence of $a^{\rm fr}$ on $\alpha$ and $\dot{m}$ in equation
(\ref{afr})), and more binaries (especially with periods $<5$ years)
will be already detached from their discs and probably less suitable for
multimessenger observations.
In an $\alpha$-disc, the viscosity is much larger in the radiation dominated zone,
and, even though there is no self consistent solution to the binary-disc evolution,
it is likely that the decoupling would happen at smaller separations. In any
case, $t_\nu$ is surely much smaller in the radiation dominated zone for an
$\alpha$-disc, and the disc 'freezing' radius would certainly be smaller, allowing
 binaries to be in touch with their circumbinary discs for longer time and at
smaller separation. In this respect, the $\beta$-disc assumption used here
is likely to be conservative.
The impact of the chosen MBH-host relation is also mild, affecting the PTA--MBHB
population by a factor of two-three at most. Also the binary eccentricity is not
very important in terms of the population itself. In our simple model, in fact, we
assume the disc migration to be independent on the binary eccentricity. A population
of circular binaries will only decouple at mildly smaller values of $a$, slightly
increasing the population of coupled systems. However, in this case, variability
related to periodic streams flowing through the gap would be highly reduced
\citep{roedig11}, making probably difficult to recognize these systems through 
periodicity studies.

As a note of caution, we remind the reader that our models assume that {\it all} merging 
MBHBs are active in their last evolutionary stage. Under such condition (and the 
assumption that systems are accreting at $\dot{m}=0.3$), the source flux distribution
shown in figure \ref{pop} accounts for $\approx 30\%$ of the soft-X AGN  
Log$N$-Log$S$ as observed by ROSAT \citep{ROSAT99}, meaning that one bright X-ray AGN
out of three is indeed a massive binary. This, although not inconsistent with any
observation, is, of course, a strong statement, that can be relaxed if $\dot{m}<0.3$,
if a certain fraction of our sources is indeed obscured, or simply if only a fraction
${\cal F}$ of MBHBs in indeed active. In this latter case, it is sufficient to rescale
our predictions by a factor ${\cal F}$, accounting for the fraction of active binaries. 
Similarly, we assumed that all MBHBs overcome the 'last parsec problem' \citep{milo01}; which,
in the light of several recent results about both gas and stellar driven MBHB hardening
\citep{Escala2005,Dotti07,Jorge09,Colpi09,Khan11,Preto11}, we consider a reasonable 
assumption.   
We will see in the following that even for ${\cal F}\lesssim 0.1$, the number of 
predicted sources is sizable, making multimessenger astronomy with PTA sources an 
appealing prospect, worth of deeper investigations.
 
\section{Dynamical modelling of the prototypical PTA-MBHB}
Having studied the MBHB population relevant to our investigation, we model in this section
the dynamics of the prototypical PTA-MBHB surrounded by a circumbinary thin accretion disc. 
We use a syncretic approach, combining analytical accretion disc models to high resolution 
SPH simulations. The former provide the global set-up of the system, the latter are necessary 
to gather a better insight of the dynamics of the torqued material streaming through the inner edge
of the circumbinary disc.

\subsection{Standard model for a PTA binary}
\label{subsec:standardmodel}
We model a MBHB of a total mass $M = M_1+M_2$
surrounded by a thin Keplerian gaseous disc of mass $M_d$. Although
Newtonian simulations are in principle scale-free, the expected
properties of the circumbinary disc depend on the assumed mass and semimajor axis of the binary. We
therefore need to work out an {\it a priori} scaling of the SPH simulation.
We set primary mass $M_1 = 2.6\times 10^8 M_{\odot}$, mass ratio $q=M_2/M_1 = 0.35$ and semimajor axis
$a_0=0.012$ pc, consistent with the typical range of source parameters found in the previous section
(see figure \ref{pop}). Following \cite{roedig11}, we set the the initial binary eccentricity
to a value $e_0=0.6$. The binary has angular velocity $\Omega_0 = (GM/a_0^3)^{1/2}$
resulting in a period $P=f_0^{-1}=2 \pi / \Omega_0\approx 6$ years ($f_0$ is the orbital
frequency). The binary is surrounded by a corotating coplanar thin disc, a configuration
expected in the late stage of a relatively 'wet' galaxy merger, as shown by several
SPH simulations \citep{Mayer07,Dotti2009b}.
A truncated circumbinary $\beta$-disc has mass \citep{Haiman2009}
\begin{equation}
M_{\rm d}=8.63\times10^4{\msun}\,\alpha_{0.3}^{-4/5}\left(\frac{\dot{m}_{0.3}}{\epsilon_{0.1}}\right)^{7/10} M_8^{11/5}(R_{\rm out}^{5/4}-R_{\rm in}^{5/4}),
\end{equation}
where two limiting radii of the disc, $R_{\rm in}$ and
$R_{\rm out}$, are expressed in units of $ 10^3R_{\rm S}$
and, in our simulation, correspond to $R_{\rm in}=2a_0$ and $R_{\rm out}=10a_0$, respectively.
Plugging in our default disc model ($\alpha_{0.3}=\dot{m}_{0.3}=\epsilon_{0.1}=1$) and the
binary parameters assumed above, we obtain $R_{\rm in}\approx 0.6$,  $R_{\rm out} \approx 3$ and
$M_d\approx 5\times 10^6\msun \approx 1.5\times 10^{-2} M$.
Comparing the viscous timescale of the binary to the GW driven orbital
decay (equation (\ref{tnu}) and (\ref{tgw}), in Section 2.2), we find that the MBHB
is still attached to the circumbinary disc, and we can study the dynamics in the
Newtonian approximation, neglecting relativistic effects.

\subsection{Numerical realisation}
\label{subsec:numerical}
We use the SPH code {\sc Gadget}-2 \citep{Springel05} as done in \cite{roedig11},
we refer the reader to that paper for details and only outline the essentials here.
As initial condition, we use the binary outlined in the previous section surrounded by
an $8$ million particle circumbinary disc, and let it relax over $9$ orbits.
The dynamics of the MBHs, which are modelled as Newtonian point masses, is followed with a fixed
time-step, equal to 0.01 $\Omega_0^{-1}$.
The gas is allowed to cool on a time scale which is proportional to the local dynamical
time of the disc thus we set $\beta = t_{\rm cool}/t_{\rm dyn} = 10$. In order to confine the
gas in the central cavity to a thin geometry, we  assume that the small amount of gas
present in the inner cavity ($r\lsim 1.75a$) is isothermal, with an internal energy per unit mass $u
\approx 0.14 (GM/R)$. The softening for the particles is adaptive with a minimum of 
0.001 $a_0$, while the MBHs are not softened \footnote{Note that this does not introduce any
artificial effect in the computation of the gravitational forces, since the MBHs are modelled 
as sink particles with sink radius larger than the typical softening of the gas particles.}.
The sink radius, the radius below which any passing particles is counted "accreted"  
\citep{Bate95}, is set to $0.005\, a_0$, which corresponds to $\approx 6\, R_S$ for the secondary MBH. 
The necessity of such a high resolution is given by the high mass ratio between
holes and discs. Furthermore, we need to be able to resolve sub-Eddington accretion rates
onto the black holes, which is $\dot{M}_{\rm Edd} = 8 \msun/$yr, thus we need to
have single particles that are some factor smaller than this. In our choice of units,
the mass of a particle is $M_{SPH}= 0.79\, M_{\odot} $.
The aspect ratio of the disc is  $h/r\sim 0.04$ with a mass ratio between disc and holes
$M_{\rm d}=1.58\times 10^{-2} M$.

\subsection{Variable inflows and minidisc formation}
\label{subsec:lightcurves}
\begin{figure}
\centering
\hspace*{-0.1cm}
\includegraphics[width=\linewidth]{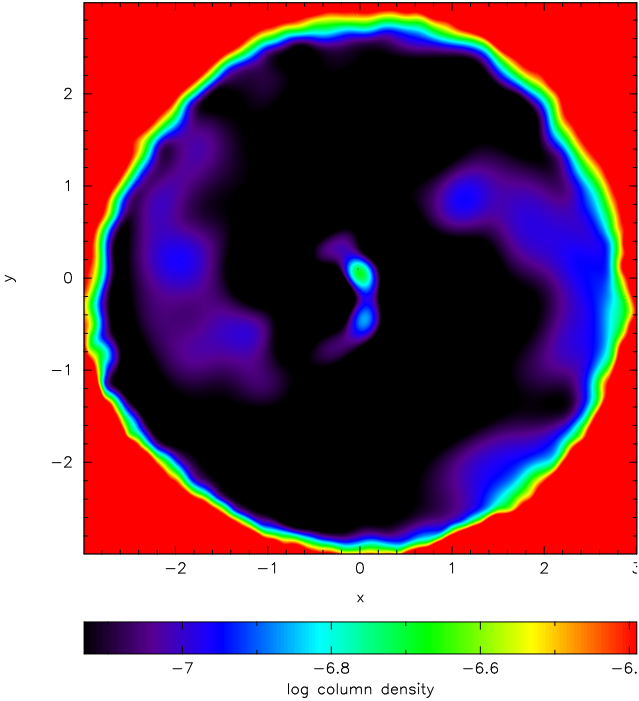}
\caption{Logarithmic column density of the cavity, the MBHs are inserted as small black dots, the axes are
in units of the semimajor axis $a_0$}
\label{fig:minidisc}
\end{figure}

\begin{figure}
\includegraphics[width=\linewidth]{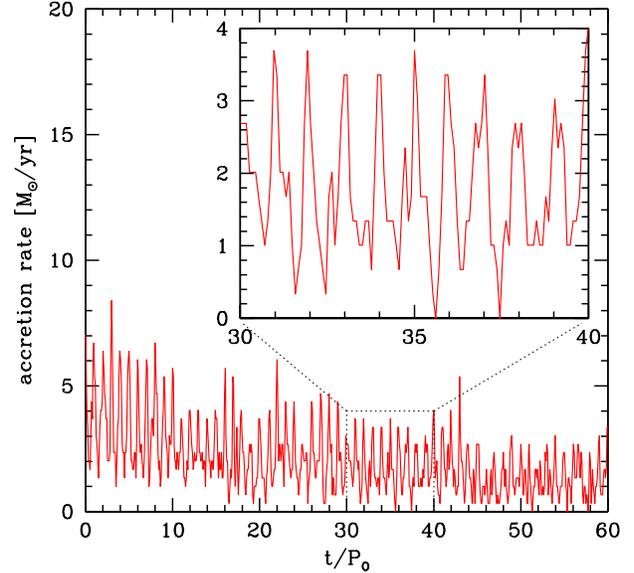}
\caption{Numerical accretion rate as a function of time (see text for details).
The periodicity is remarkable, as also highlighted in the upper-right zoom-in.}
\label{accrate}
\end{figure}

Gravitational torques, acting on the inner edge of the circumbinary
disc, cause periodic inflows of mass that are triggered by the
secondary pulling gas out of the disc edge at each apo-apsis passage.
We follow the streams down to the sink radii of the two black holes and
monitor the particles that are {\it swallowed} by the MBHs and bin
them according to the timestep when they were swallowed.
An example of the forming streaming structure is given in figure~\ref{fig:minidisc},
where we show the linear column-density of the gas in the gap
\footnote{fig.~\ref{fig:minidisc} made using {\sc SPLASH}
  \citep{Price2007}.}. The inflows are well resolved, and the gas bound
to each MBH is visible to form a sense of minidiscs. Their radial size is
about $0.05\,a_0$ and their average density $\rho= 2\times
10^{-15}$g cm$^{-3}$, which gives a total mass of the minidisc $m_{{\rm
    md}_1}\sim 3 \msun$ for the primary and $m_{{\rm md}_2}\sim 0.5
\msun$ for the secondary. The detailed structure of the inflows and
their physical properties strongly depend on the disc thermodynamics;
moreover, their behaviour close to the MBHs is likely to be affected by
the comparable size of the sink radius, the smoothing length and the
physical size of the minidiscs. Nonetheless, we can consider the rate
at which particles cross the sink radii of the two black holes (what
we call {\it numerical} accretion rate) as a good proxy to the rate at
which material is fed from the outside circumbinary disc to the inner
region. Such a rate is shown in figure~\ref{accrate} over sixty orbital
periods.  The orbital periodicity is quite striking, with a
maximum-to-minimum ratio larger than a factor of three. 
The maximum accretion rate declines from $\dot{m}\sim1$ to $\dot{m}\sim0.4$,
as a result of the relaxation of the disc from its initial condition which is
slightly out of equilibrium. We notice that the inflows
occur on the MBHB orbital period, and are triggered by the secondary
pulling material off the inner edge of the circumbinary disc at its
apo-apsis passages.  Their nature is therefore primarily
gravitational, and should be weakly dependent on the exact thermodynamics of
the disc. Our simulation therefore suggests that gravitational torques
are effective in feeding periodically the central gap at a rate which
is a substantial fraction of the Eddington rate.
Our results are in agreement with similar full GR hydro
simulations studying the evolution of MBHBs at closer separations
\citep{farris11}, in which similar streams feeding the two MBHs are
observed down to $\approx 10 M$ ($\approx 3\times 10^{-4}$ pc for our
MBHs). 

\subsection{Analytical modelling of the minidiscs}
\label{subsec:lightcurves}
\begin{figure}
\includegraphics[width=\linewidth]{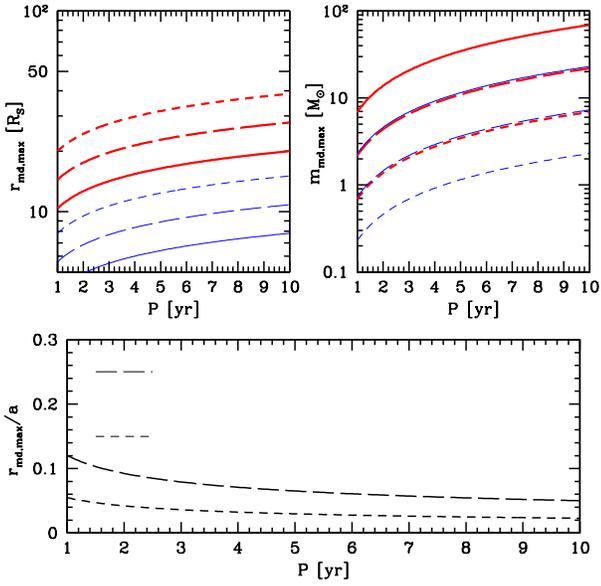}
\caption{Characteristic properties of $\alpha$-minidiscs. {\it Upper left panel}:
maximum minidisc radius $r_{\rm md,max}$ allowing for $t_{\rm acc, \alpha}<P$,
as a function of $P$. Thick-red curves are for $\dot{m}=0.3$ and $\alpha=0.3$
and thin-blue curves are for $\dot{m}=0.1$ $\alpha=0.1$. Short--dashed, long--dashed
and solid curves are for a MBH mass of $10^8$, $10^{8.5}$ and $10^9\msun$ respectively.
 {\it Upper left panel}: maximum mass of the minidiscs enclosed in $r_{\rm md,max}$
as a function of $P$, line and color style as in the upper left panel.
{\it Lower panel}: maximum size of the minidiscs relative to $a$ as a
function of $P$, compared to the Roche-lobes of the two MBHs (horizontal ticks).
Here we assume our default binary; long--dashed lines are for $M_1$, short--dashed
are for $M_2$.}
\label{minidiscs}
\end{figure}

As stated in the previous section, the detailed behaviour of the inflows close to the
MBHs is likely to be affected by the comparable size of the sink radius, the smoothing
length and the physical size of the minidiscs. We therefore decide, for the purpose of
the following discussion, to build a simple analytic parametric model for the
minidiscs. We assume that the infalling material is fed from the outer disc at some
given rate $\dot{m}$. To simplify the notation, we do not distinguish between the two
MBHs; all the quantities appearing in the equations below, are intended to be specific
of the MBH under consideration{\footnote{For example, $M_8=M_i/10^8$M$_\odot$ is the
    mass of the $i$-th MBH in units of $10^8$ solar masses,
    $\alpha_{0.3}=\alpha_i/0.3$ is the viscosity parameter of the $i$-th disc in units
    of 0.3, and so on.}}, and the results apply to both MBHs (and both attached
minidiscs).  The infalling material is captured in the Roche lobe of the $i$-th MBH
with an average specific angular momentum ${\cal J}$ with respect to that MBH.  This
is converted to a characteristic size of a Keplerian orbit around the MBH simply as
$r_{\rm md}={\cal J}^2/GM$. If $r_{\rm md}<3R_{S}$ the inflow is radial and accretion
proceeds in a Bondi-like fashion. Otherwise, the infalling material settles at a
characteristic radius $r_{\rm md}$ around the MBH and dissipates its angular momentum
through viscous processes, diffusing inwards and eventually being accreted by the two
holes \citep[a process similar to the debris fallback in tidal disruption induced
  accretion,][]{rees1988}.  The time needed for an annulus of material at radius $r$
to be accreted is
\begin{equation}
t_{\rm acc, \alpha}=0.31 \,{\rm yr}\, \alpha_{0.3}^{-1}\left(\frac{\dot{m}_{0.3}}{\epsilon_{0.1}}\right)^{-2}M_8\, r_1^{7/2}
\label{taccalpha}
\end{equation}
\begin{equation}
t_{\rm acc, \beta}=683 \,{\rm yr}\, \alpha_{0.3}^{-4/5}\left(\frac{\dot{m}_{0.3}}{\epsilon_{0.1}}\right)^{-2/5}M_8^{6/5} r_1^{7/5}
\label{taccbeta}
\end{equation}
for $\alpha$ and $\beta$ discs respectively{\footnote{To sketch the situation,
we describe the minidiscs as small steady accretion discs, even though
the condition of stationarity of the system are
not met because of the periodicity of the infalling streaming.}}. Here $r_1$ is the size of the
minidisc in units of $10R_{S}$. If $t_{\rm acc}$ is longer then the binary
orbital period $P$, then a pair of persistent, periodically fed minidiscs is formed 
around the two MBH. If, conversely, $t_{\rm acc}$ is shorter then $P$, then the streaming 
periodicity is reflected in episodes of periodic accretion. For $\alpha$-discs, this happens for
\begin{equation}
r_{\rm md,max}<20 \,R_{S}\, \alpha_{0.3}^{2/7}\left(\frac{\dot{m}_{0.3}}{\epsilon_{0.1}}\right)^{4/7}M_8^{-2/7}P_{\rm yr}^{2/7}
\end{equation}
($P_{\rm yr}$ is the MBHB period in years), while it never happens for
$\beta$-discs. Figure \ref{minidiscs} shows the maximum
radius $r_{\rm md,max}$ and the corresponding maximum mass $m_{\rm md,max}$ of an
$\alpha$-minidisc. For average feeding rates of  $\dot{m}\approx 0.3$
(consistent with the output of our SPH simulation), maximum disc sizes are in the range
10-40 $R_S$.

In our specific simulation, the average angular momentum of the
particles captured by the MBHs within their respective Roche--lobes implies
$r_{\rm md}\approx 10-20 R_S$, meaning that the accretion process must happen
through viscous inspiral rather
than through radial infall, disfavoring the Bondi-like accretion scenario.
Moreover, $r_{\rm md}<r_{\rm md,max}$ for both MBHs; for $\alpha$-discs, this suggests MBHB
period-related periodicity in the accretion and thus in the emitted luminosity.

\section{Emitted spectrum and electromagnetic signatures}
\begin{figure}
\includegraphics[width=\linewidth]{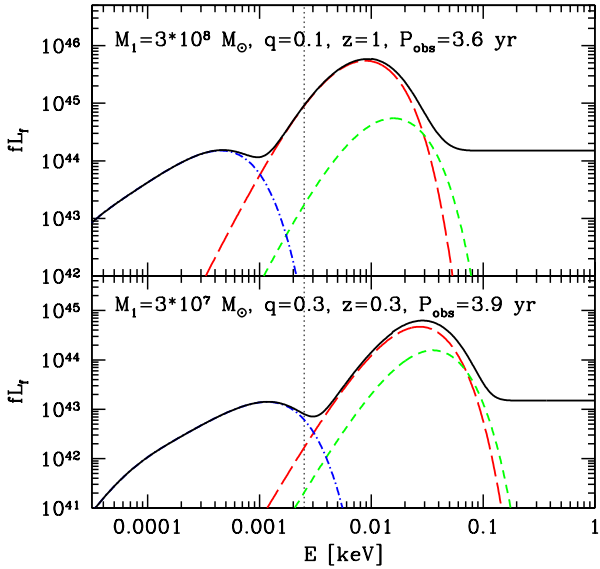}
\caption{Continuum components of the emitted spectra. In each panel we plot the
thermal emission of the circumbinary disc (blue dot--dashed line), and of the
two inner minidiscs (long--dashed red line for the primary, short--dashed green
line for the secondary); depending on the nature of the inflowing streams, these latter
components may or may not be present. The black solid line is the resulting total continuum,
where a flat X-ray emission from a putative hot corona has been added (see text). The
vertical dashed line is the energy center of a typical green optical filter. The two
panels refer to two fiducial systems with parameters labelled in figure. In the upper
panel, each minidisc has size $r_{\rm md}=25R_S$ of the correspondent MBH; while in the
lower panel, the size of the minidiscs is $r_{\rm md}=15R_S$ only. We assumed our 
default accretion model ($\alpha=0.3$, $\dot{m}=0.3$, $\epsilon=0.1$).}
\label{spectrum}
\end{figure}

In optically thick disc theory, each disc annulus emits
blackbody radiation corresponding to its particular temperature;
for a standard optically thick geometrically thin disc, this is
\citep{frank2002}
\begin{equation}
T(r)=\left[\frac{3GM\dot{M}f(r)}{8\pi r^3 \sigma_{\rm sb}} \right]^{1/4},
\end{equation}
where $f(r)=(1-3R_s/r)^{1/2}$ (we assume a Schwarzschild black hole), and $\sigma_{\rm sb}$
is the Stephan-Boltzmann constant. The emitted luminosity at frequency $\nu$ is
given by the integral over the disc extension of the blackbody radiation emitted
by each annulus, i.e.
\begin{equation}
L_{\nu}=\frac{32\pi h_P \nu^3}{c^2}\int_{r_{\rm in}}^{r_{\rm out}}\frac{rdr}{e^{h_P\nu/kT(r)}-1},
\label{lnu}
\end{equation}
where $r_{\rm in}$ and ${r_{\rm out}}$ are the inner and outer edges of the disc
respectively, $h_P$ is the Planck constant, and $k$ is the Boltzmann constant. In our
model, each of the discs (the circumbinary plus the two minidiscs) emits a thermal
component according to equation (\ref{lnu}).  We therefore infer the presence of three
distinctive continuum components: (i) a thermal component peaked in the optical/IR
emitted by the circumbinary discs; (ii) two thermal component peaked in the UV
associated to the inner minidiscs; (iii) two X-ray powerlaw associated to the tenuous
hot electron plasma \citep[corona,][]{galeev1979} surrounding the inner
minidiscs. These latter components are generated by the inverse Compton scattering of
UV photons emitted by the inner radii of the minidiscs against the diffuse hot plasma
of electrons embedding the inner part of the discs.  The up-scattered photon spectrum
can be modelled as a power-law spanning the soft/hard-X domain.  For the qualitative
nature of our discussion, it is sufficient to consider a flat spectrum in $\nu L_\nu$
\citep{haardt1993}, normalized to give $L_{0.5-10 {\rm keV}}=0.03L_{\rm bol}$
\citep{Lusso10}.  The general features of the predicted continuum are shown in figure
\ref{spectrum} for two selected systems. The presence of a gap is reflected by the
characteristic double bump; the one in the IR is due to the circumbinary disc, while
the one in the UV is produced by the two inner minidiscs.

On top of the continuum, several sets of emission lines are also expected, as in
common AGNs. In particular: (i) optical broad emission lines caused by the
photoionization of the tenuous infalling material and the outer circumbinary disc
($r>0.01$pc) by the inner ionizing UV source (the minidiscs, $r<10^{-3}$pc); (ii) two
6.4keV fluorescence iron broad emission lines (Fe K$\alpha$ lines) produced by the
reflection of the up-scattered corona X-ray photons on the surface of the inner
accretion minidiscs at only few Schwarzschild radii \citep{Matt91}. Moreover, several
other spectral features may be present: (i) some weak optical/UV emission associated
to the streaming material \citep{Bogdanovic2009,Dotti2009}; (ii) X-ray hot spots
related to the instreaming material shocking onto the outer edge of the inner
minidiscs (if $t_{\rm acc}>P$, see Section 5.1).

\subsection{Characteristic signatures}
Depending on the strength and angular momentum of the inflows, the features outlined
above may or may not be present. We identify three different scenario:
\begin{enumerate}
\item {\bf $r_{\rm md}<3R_S$}. The streams flow radially onto the two MBHs. The
  radiation emission of the associated Bondi-like accretion is likely to be extremely
  inefficient.  No ionizing UV continuum and, consequently, no optical broad lines are
  present.  In this scenario, no signatures of the existence of an accreting system
  may be detectable. The only guaranteed component is the continuum associated to the
  circumbinary disc thermal emission. If, however, the spread in specific angular
  momenta of the accreted matter is large compared to the angular momentum of the last
  stable orbit ($\Delta{\cal J}>R_Sc$), then a scenario analog to the one proposed by
  \cite{illarionov2001} for wind-fed high mass X-ray binaries may take place. The
  inflowing streams collide in a caustic at few $R_S$ and are promptly accreted in a
  free-fall time, forming a small scale, inviscid, radiatively efficient accretion
  disc \citep{beloborodov2001,zalamea2009}. In such models, the accreted
  matter/luminosity conversion efficiency ranges from $0.03$ for Schwarzschild MBHs,
  up to 0.1 for maximally spinning MBHs. The MBHB would therefore appear as a
  periodically variable, luminous X-ray source.

\item {\bf $3R_S<r_{\rm md}<R_{\rm max}$, $\alpha$-discs only}. The streams form
  standard minidiscs around the two MBHs. Their typical viscous time is shorter than
  the binary orbital period; therefore, accretion onto the two MBHs can be directly
  linked to the periodic streams. In this scenario, efficient thermal emission comes
  from both the circumbinary disc and the two minidiscs. While emission from the
  circumbinary disc will be relatively steady, strong periodicity in the UV/soft-X
  will reflect the periodic accretion related to the minidiscs. The resulting periodic
  ionizing photon flux, will cause optical broad line strength to fluctuate by a
  factor larger than two over the orbital period. However, identification of such
  optical variability may be challenging, since it would require an all sky periodic
  spectroscopic monitoring. Large photometric surveys like the LSST \citep{LSST2009},
  will be sensitive to the luminosity of the optical continuum, which, depending on
  the characteristics of the system, may be dominated by the steady circumbinary disc
  (see figure \ref{spectrum}). Nonetheless, depending on the amount of reddening,
  color selection may identify candidates showing a distinctive optical bump across
  the different wavelength bands.
Note that in such close systems, the broad line emission region is likely related to
the circumbinary disc; the typical size of the broad line emission region in AGN is
$\sim 0.01-0.1$pc, much larger than the size of the minidiscs, and correlates with the
AGN luminosity \citep{kaspi2005}.  Detection of separate sets of broad optical
emission lines related to accretion onto the individual MBHs, or the detection of a
broad emission line shifted with respect to the narrow emission line associated to the
host \citep[see, e.g.][]{decarli2010} are therefore unlikely in this case. Periodic
soft-to-hard X-ray emission in response to the periodic continuum may be associated to
the hot corona, and variable relativistic Fe K$\alpha$ lines may also be present. This
is the scenario suggested by our SPH simulation.
\item {\bf $r_{\rm md}>R_{\rm max}$ ($\alpha$-discs), or $r_{\rm md}>3R_S$
  ($\beta$-discs)}.  The streams cannot be swallowed by the MBHs within an orbital
  period. Steady thin accretion minidiscs form around the two MBHs. The accretion rate
  may be still quite variable, but redistribution of angular momentum within the
  minidiscs makes impossible to link the accretion variability to the periodicity of
  the streams. Two superposed relativistic Fe K$\alpha$ lines associated to the
  minidiscs are likely to be observable. Moreover, shocks created by the collision of
  the instreaming material and the minidiscs at their outer edge may result in X-ray
  hotspots varying on the orbital period. The luminosity and electromagnetic frequency
  at which these regions irradiate depend on the relative velocity ($v_{\rm rel}$) of
  the streams with respect to the minidiscs. We can estimate $v_{\rm rel}$ for $M_2$,
  when the secondary is at the apocenter, assuming that its minidisc fills the whole
  Roche lobe. In this case, the emission is expected to peak in the hard X-ray, at
  about $\sim v_{\rm rel}^2 \mu_p/k \gsim 10$ keV ($\mu_p$ is the mean molecular
  weight of the plasma). The luminosity depends on the time-scale $\tau_{\rm shock}$
  over which the shock irradiates its internal energy. Assuming $\tau_{\rm shock}$ to
  be of the order of the orbital period of the gas at the edge of the minidisc
  ($P_{\rm disc}$), we obtain $L_{\rm shock} \approx \dot{m} v_{\rm rel}^2 P/P_{\rm
    disc} \gsim 10^{-3} L_{\rm Edd}$. Smaller $\tau_{\rm shock}$ and larger $v_{\rm
    rel}$ result in larger luminosities, but increasing $v_{\rm rel}$ shifts the peak
  of the emitted luminosity at considerably higher frequencies. Note that low values
  of $v_{\rm rel}$ are the most favorable conditions to efficiently perturb and bind
  large amounts of gas.
\end{enumerate}

In the following we will focus on two distinctive signatures in the X-ray domain.
\begin{itemize}
\item {\it Periodic X-ray emission related to the periodicity of the minidisc fed
through the gap}. The outer circumbinary disc is relatively
stable, and optical variability related to the streams may be overwhelmed by the outer
disc continuum. Periodic variability,
related to (i) periodic accretion of the minidiscs, (ii) the associated varying flux
of UV photons up-scattered in the corona, (iii) X-ray hotspots created by the periodic
streams colliding with the minidiscs, may instead be easily detectable in UV and in X-ray,
\item {\it Double relativistic fluorescence Fe K$\alpha$ lines}. Such lines are a
  common feature in AGNs \citep{nandra07,delacalle10}, and are produced at only few
  Schwarzschild radii. The line profile strongly depends on the location of the last
  stable orbit orbit around the MBH (which depends on the spin magnitude) and on the
  disc inclination with respect to the observer. If accretion is efficient on two MBHs
  with different spin parameters (magnitude and/or orientation), a distinctive 'double
  Fe K$\alpha$ line' feature may be observable in the hard X spectrum of the source.
\end{itemize}

In the two following sections we will consider these two possibilities separately,
proposing possible strategies for combining X-ray and PTA observations in the final
discussion.

\section{X-ray periodic variability}
\label{sec:EM}

\subsection{Sub-population of observable periodic sources}

\begin{figure*}
\begin{tabular}{cc}
\includegraphics[scale=0.40,clip=true]{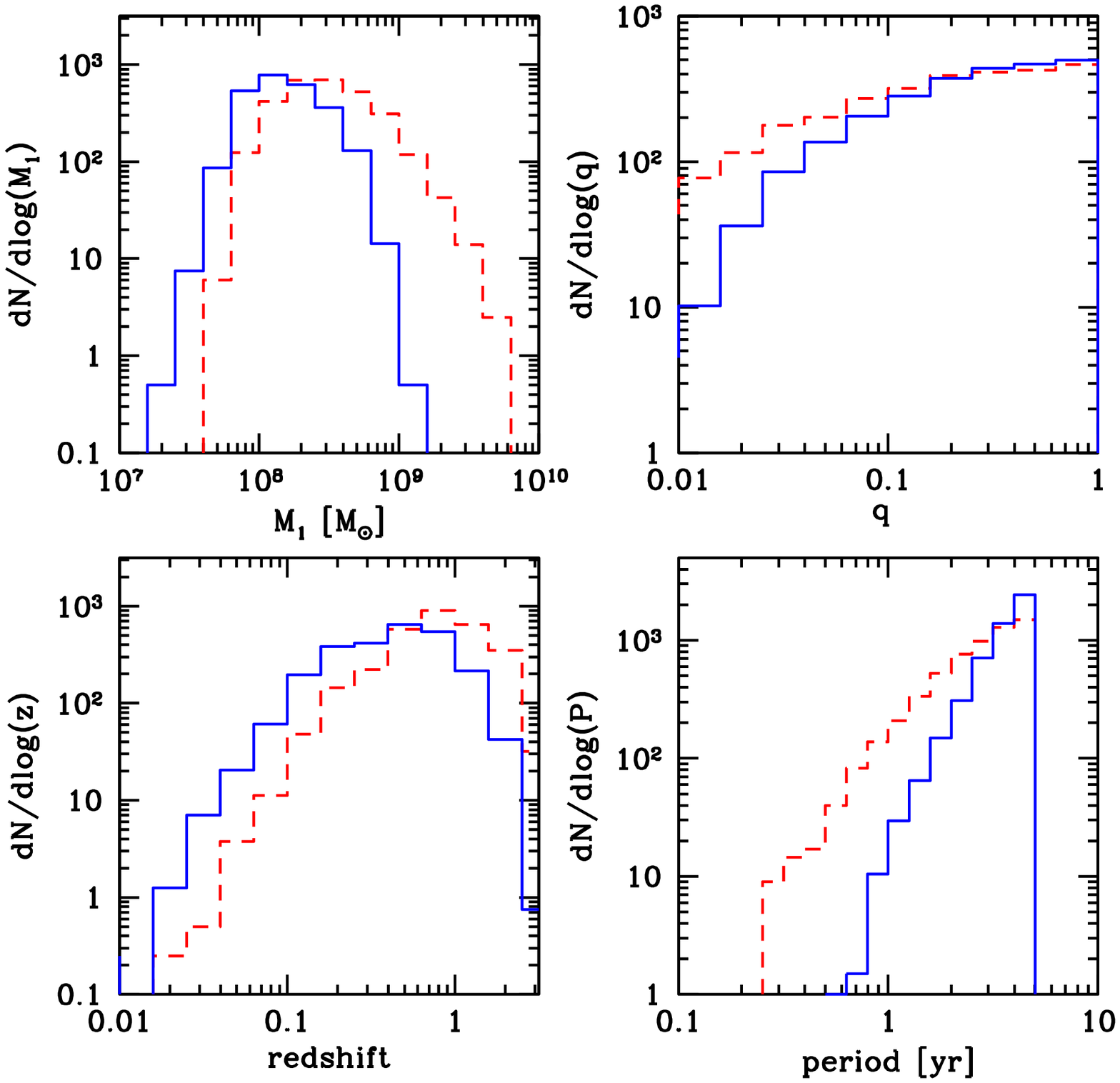}&
\includegraphics[scale=0.40,clip=true]{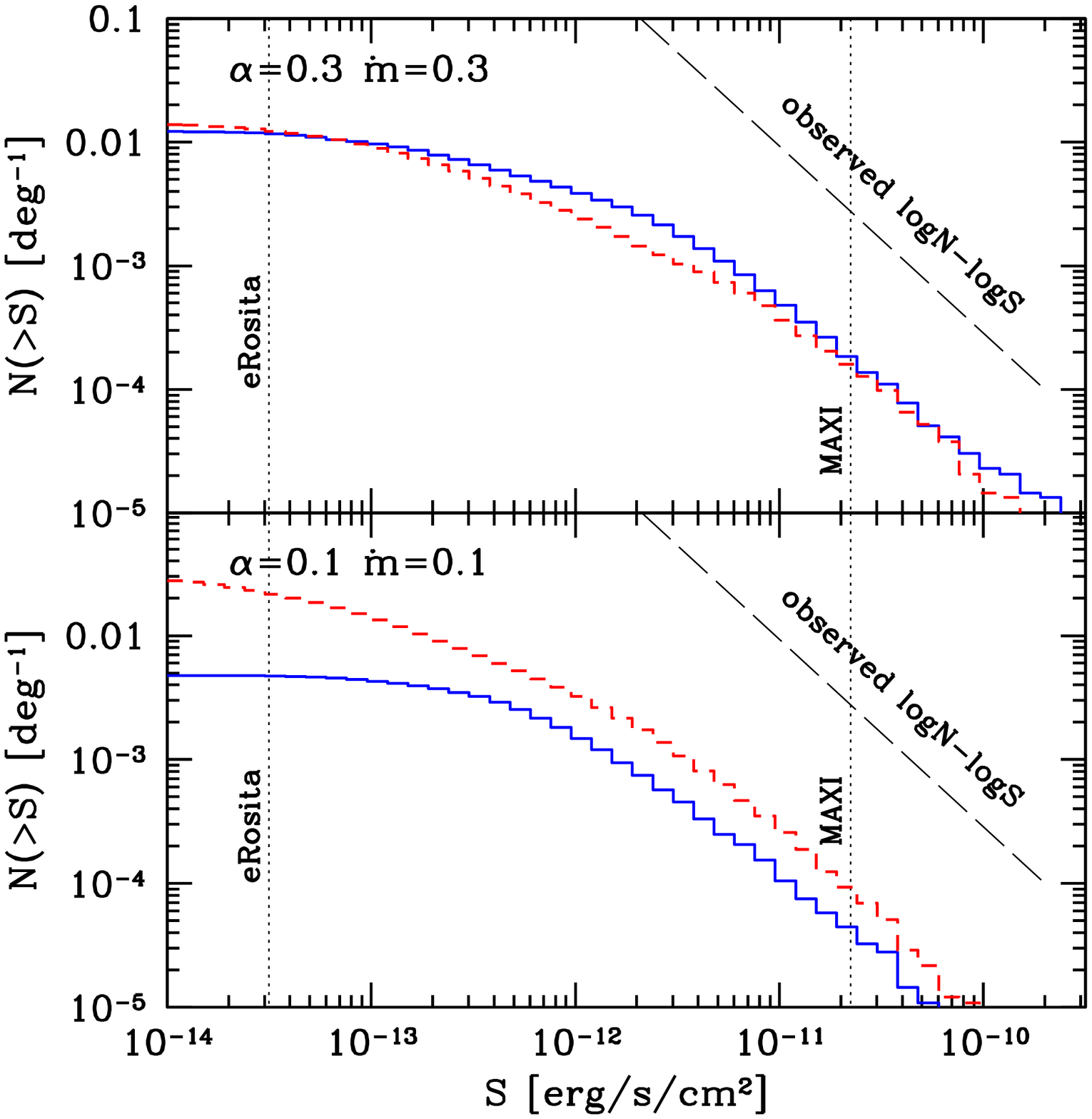}\\
\end{tabular}
\caption{{\it Left plot}: properties of the PTA-MBHB population observable through
  periodicity, contributing at a level of 0.1ns or more to the PTA signal in the
  $3\times10^{-9}-10^{-6}$Hz frequency window in our default model, averaged over 100
  Montecarlo realisations. Systems with orbital periods $<5$ years only were
  selected. From the top left to the bottom right, we plot the primary mass $M_1$,
  mass ratio $q$, redshift and period distributions.  {\it Right plot}: periodic MBHB
  contribution to the X-ray Log$N$-Log$S$ function, assuming a bolometric correction
  of $3\%$ (0.5-10 keV). The top panel is for our default model; the bottom panel is
  for a less optimal model ($\dot{m}=0.1$) for which MBHB-disc decoupling occurs
  earlier. Vertical dotted lines depict the flux limit for a single eRosita pointing
  and for the one-month sum pointings of MAXI. Dashed thin line depicts the upper end
  of the AGN Log$N$-Log$S$ as observed by ROSAT \citep{ROSAT99}. In both plots,
  linestyle as in figure \ref{pop}.}
\label{periodic}
\end{figure*}

Being interested in variability related to the binary orbits, only a subset of the
population shown in figure \ref{pop}, having reasonably short periods will be suitable
targets. We therefore limit our study to the systems with $P<5$yr. The resulting
population, represented in the left plot of figure \ref{periodic} for our default
model, retains the mass, mass-ratio and redshift distribution of the parent one.
Detached systems obviously have a period distributions extending to $P\ll1$yr, while
binaries still attached to their discs (relevant to our investigation) are abundant at
$P>1$yr. Assuming that a fraction of $3\%$ of the bolometric luminosity is emitted in
the X-ray {\footnote{This picture is appropriate for hot corona reprocessing. Inviscid
    minidiscs have matter/luminosity efficiency conversions similar to standard
    accretion discs ($\epsilon\sim 0.03-0.1$), and even though detailed calculation of
    the spectrum are not available, we may expect comparable X-ray luminosities. If
    the minidiscs are instead steady, then X-ray luminosity emitted by shock induced
    hot spots may be more than an order of magnitude smaller ($10^{-3}$L$_{\rm Edd}$,
    compared to $3\%$ of 0.3L$_{\rm Edd}$ for corona reprocessing in our default
    model).}},
we can construct the Log$N$-Log$S$ (i.e., the cumulative number of sources having a
measured $0.5-10$keV flux larger than a certain value $S$, as a function of $S$) of
the emitting population.  This is shown in the right plot of figure \ref{periodic} for
both our default population (upper panel), and for an alternative model in which the
smaller $\alpha$ viscosity and accretion rate imply an earlier detachment of the
circumbinary disc. At large luminosities, such periodic binaries may add up to $5\%$
of the observed luminous X-ray sources in the sky. The currently operating X-ray all sky
monitor MAXI \citep{MAXI09} may already have detected an handful of them in its
surveys.  On the other hand, according to our default model, there might be up to 500
MBH binaries significantly contributing to the PTA signal, showing periodic
variability on a timescale of $1-5$ years, in the sensitivity reach of the upcoming
eRosita observatory \citep{eRosita10}. We stress, once again, that our models assume
that {\it all} merging MBH binaries are accreting in their last evolutionary phase
prior to coalescence, which may be in fact likely in gas rich mergers, but nonetheless
it remains an {\it ad hoc} assumption. If, for instance, only $10\%$ of the merging
systems is active, the number of sources showing emission periodicity drops to 10-50,
which is however still an interesting, sizable sample.

\subsection{Simulating observations: sampling and statistics}

\begin{figure*} 
\begin{tabular}{ccc}
\includegraphics[scale=0.23,angle=-90,clip=true]{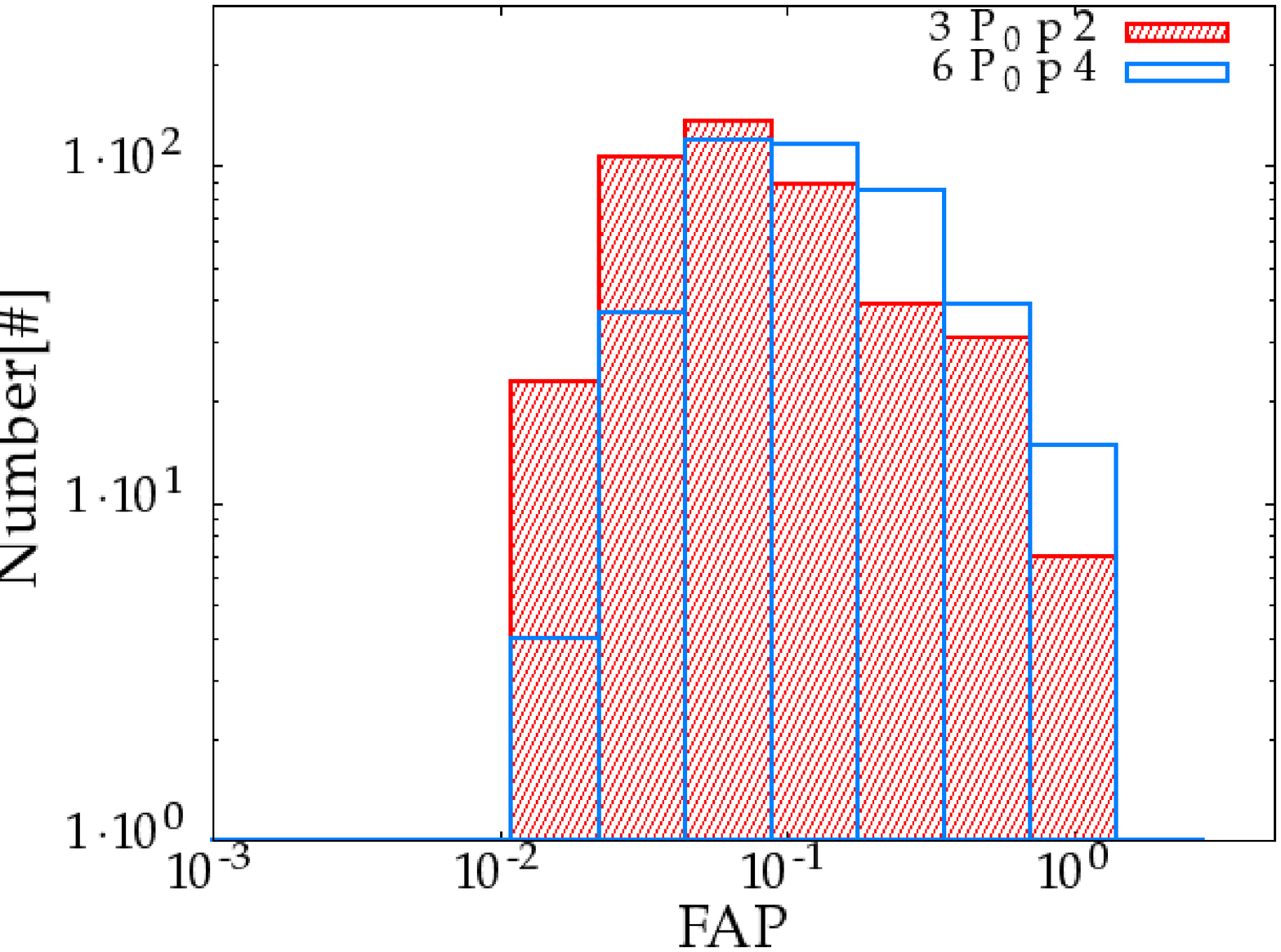}&
\includegraphics[scale=0.23,angle=-90,clip=true]{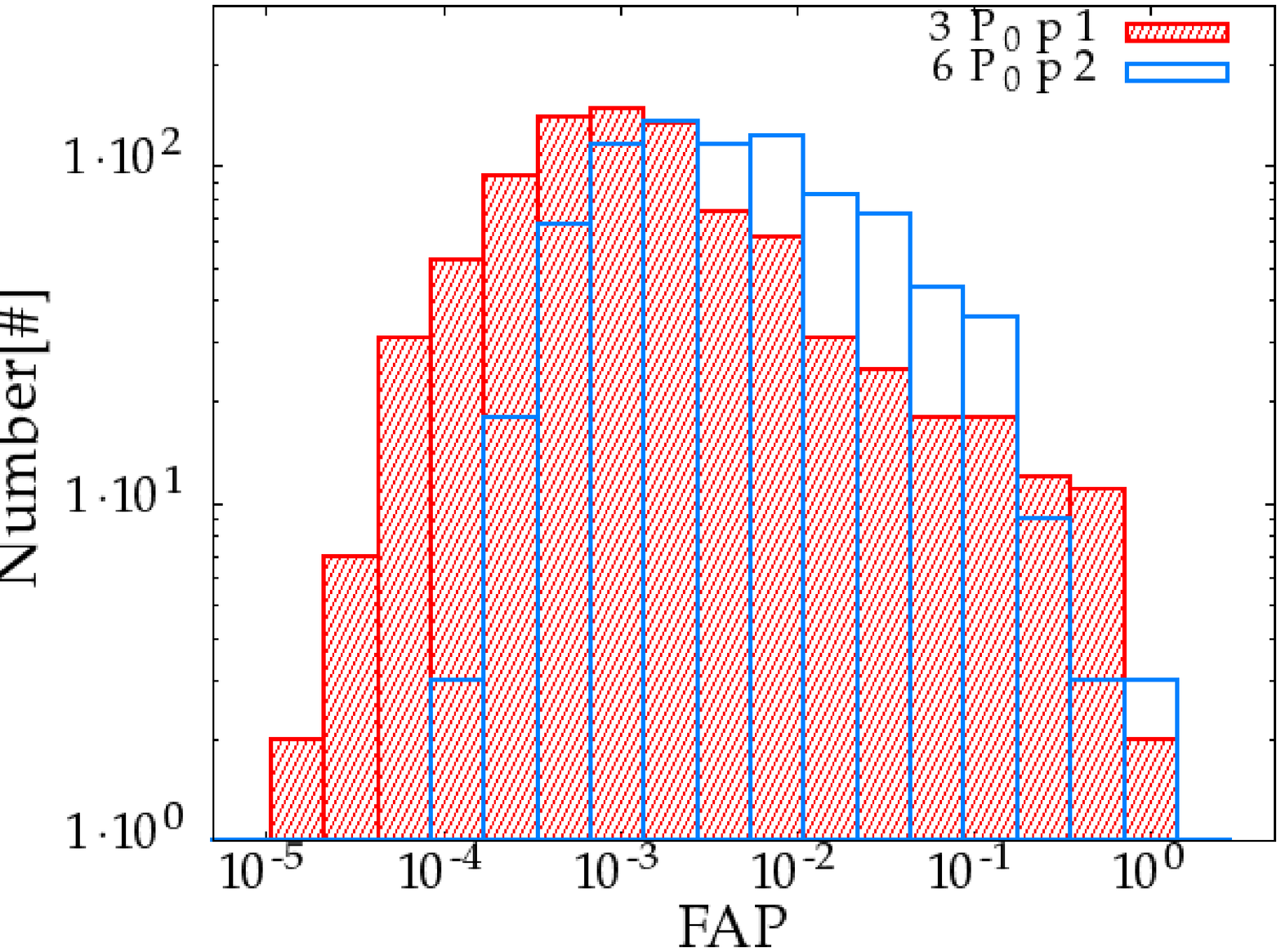}&
\includegraphics[scale=0.23,angle=-90,clip=true]{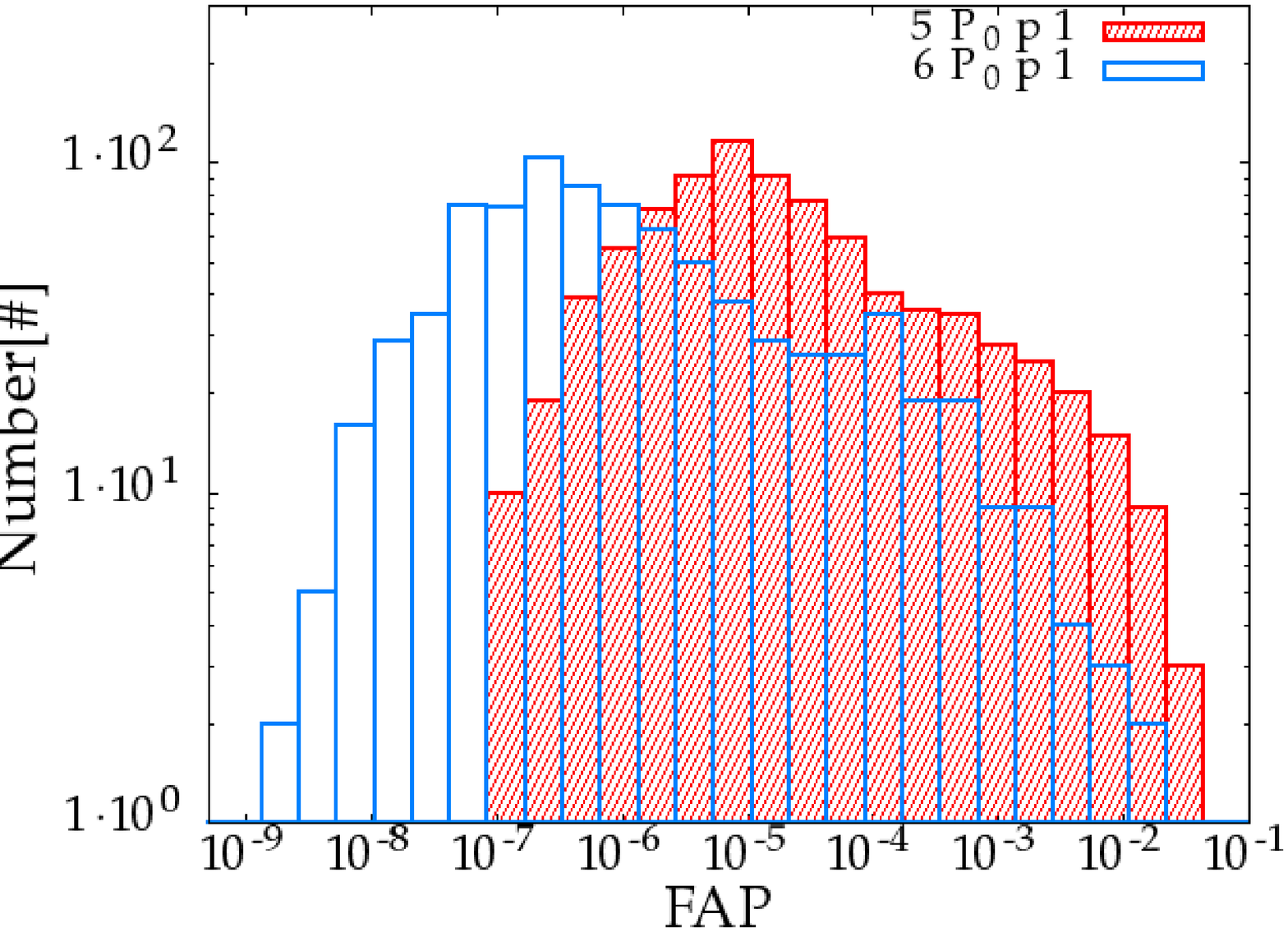}\\
\end{tabular}
\caption{Detection statistics of periodic sources. In each plot, we show histograms of
  the number of samples correctly identified within a certain False-Alarm-Probability
  (FAP). {\it Left panel}: statistics is constructed with 6 points per orbit over 3
  orbits ($3P_0p2$, red histogram), or with 3 points per orbit over 6 orbits
  ($6P_0p4$, blue histogram). {\it Central panel}: statistics is constructed with 11
  points per orbit over 3 orbits ($3P_0p1$, red histogram), or with 6 points per orbit
  over 6 orbits ($6P_0p2$, blue histogram). {\it Right panel}: statistics is
  constructed with 11 points per orbit over 5 orbits ($5P_0p1$, red histogram), or
  with 11 points per orbit over 6 orbits ($6P_0p1$, blue histogram).}
\label{periodicstat}
\end{figure*}

An X-ray luminosity above the instrument sensitivity is certainly not enough to
identify a periodic source. The lightcurve has to be reconstructed with a minimum
number of datapoints per orbit, for at least few orbits. In this subsection we carry a
detailed statistical study of the periodicity observability. The main goal is to
provide the minimum required sampling cadence an observatory must have to efficiently
identify periodicities.  Since astronomical time-series are always limited to a
certain amount of data points, we address the problem of having a very coarse sampling
rate of our lightcurve.  We assume that the X-ray lightcurve mimics the numerical
accretion rate of figure \ref{accrate}.

The complete lightcurve of our simulation consists of $L=898$ points, sampling $\sim
80$ orbits; each orbit is therefore sampled by $\sim 11$ equally spaced points. We
create subsamples of this lightcurve by selecting a random starting zero point
$\bar{p}$ and then considering all the subsequent $np$ points, where $n=1,...,N-1$,
and $p\in[1,2,3]$.  A stride of $p=1$ translates into taking $11$ points per orbit
(points $\bar{p}$, $\bar{p}+1$, $\bar{p}+2$...), $p=2$ means taking $\sim 6$ points
per orbit (points $\bar{p}$, $\bar{p}+2$, $\bar{p}+4$...) and $p=3$ means taking $\sim
4$ points per orbit (points $\bar{p}$, $\bar{p}+3$, $\bar{p}+6$...). We refer as
$6P_0p1$ to a sample covering six orbits with all points included in the sampling
(total of $N=65$ points); $3P_0p2$ refers to a string of three orbits sampled every
other point (total of $N=18$ points), etc.  The reason to consider different sampling
is to investigate the effect of the observation cadence.  Assuming a binary with,
e.g. 3 year period, $p1$ corresponds to a three month cadence, $p2$ to six months,
etc.  If the number of points covered by the typical subsample is $N$, we can draw a
maximum of $\zeta = L-N$ subsamples {\footnote{Note that the subsamples in general
    overlap with each other, therefore cannot be regarded as uncorrelated. However, a
    much longer simulation would be needed to draw a large number of uncorrelated
    samples, which would have been prohibitively time consuming.}}.  To account other
possible sources of variability \citep[AGNs are generally variable in the
  X-ray,][]{grupe2001}, the samples are then polluted with Gaussian noise $H$ with a
dispersion $\sigma=10\%$ of the maximum accretion rate/luminosity of each specific
sample. Each sub--sample is Fourier--transformed using the Normalized Lomb--Scargle
Periodogram \citep{scargle82} and the False--Alarm--Probability (FAP) of the highest
detected peak is computed.
After a cross--check that this highest peak corresponds to the fundamental frequency
of the binary $f_0=1/P_0$, the FAPs of the $f_0$ frequency are binned logarithmically
and shown in the histograms in figure \ref{periodicstat}. The left panel shows that a
minimum of $\sim 15-20$ datapoints covering at least three orbits is needed in order
to identify a decent fraction $\sim50\%$ of the sources at a $10\%$ false alarm
probability level. Increasing the number of datapoints dramatically increases
identification performances; already with $\sim 30$ points in the lightcurve we can
detect most of the sources $\sim 70\%$ at a false alarm probability level of $1\%$.

\begin{figure}
\centerline{\psfig{file=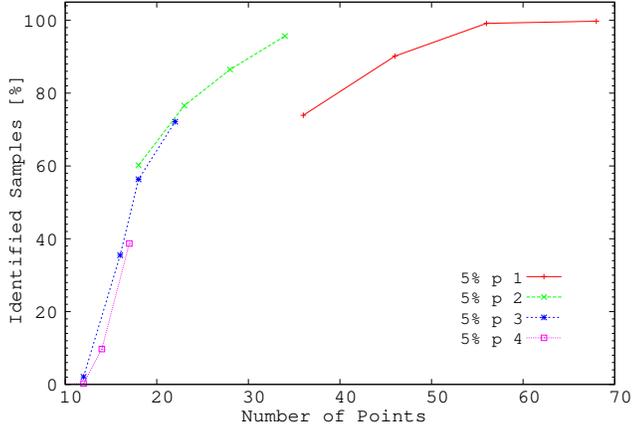,width=84.0mm}}
\caption{Percentage of correctly identified frequencies above 3 $\sigma$ within a
  fractional frequency error $<5\%$, as a function of the number of points in the
  lightcurve. For each sampling (shown in different colors, see legend in figure),
  points from the left to the right correspond to three, four, five and six full
  orbits of the observed source. A 10\% random Gaussian noise was added to each
  subsample.}
\label{fig:percentages}
\end{figure}

Additionally, it is of interest, in order to crosscheck with a possible PTA
identification of the source, to which accuracy the fundamental frequency can be
recovered from the Fourier Spectra. For an observation time $T$, the frequency
resolution bin of the observation is $1/T$. If therefore we observe a frequency $f_0$,
the relative accuracy of the observation is $\Delta f_0/f_0=(1/T)/f_0$. However, we
can oversample the spectrum to get a better constrain on the observed frequency. For
all periodograms shown, we have used an oversampling factor (ofac) of $8$. In evenly sampled
data, we expect the analysis to be only mildly dependent on this parameter: For some
choices of low ofac ($< 4$) in very short data sets, we observed the
appearance of aliasing especially if high frequency components were not suppressed
(hifac $> 1$ as defined in \citep{scargle82}). We thus concluded, especially in the
prospect of real unevenly sampled astrophysical data, that the high frequency noise
should be suppressed using very low hifac ($< 1$) and at the same time using
high oversampling (ofac $> 4$).  For the identification of any frequency we set the
requirement $|f_{\rm detected} - f_{\rm true}|/f_{\rm true}<5\%$. In figure
\ref{fig:percentages}, we show the dependence of the number of correctly identified
frequencies on the number of points available in the lightcurve, for different
samplings (i.e., for different number of points per orbital period).  We consider only
frequency peaks identified at a $3\,\sigma$ significance or more.
Note the continuity of the p2, p3 and p4 curves (six, four, and three points per
orbit, respectively), meaning that, as long as we sample three or more orbits, the
fraction of correctly identified systems depends strongly on the total number of
points in the lightcurve and only mildly on the number of points per orbit (as long as
this is larger than three). The discontinuity with respect to the p1 series tells us
that sampling six orbits with six points per orbit is much better than sampling three
orbits doubling the observation frequency. Sampling at least three orbits is a minimum
identification requirement, because other sources of uncorrelated noise may wash out
the significance of the periodicity if the statistics is too poor. In general, having
at least $\sim15-20$ points in the lightcurve, sampling three or more orbits, is a
minimum requirement for a confident identification of the periodicity.
     
\section{Double iron lines}

\subsection{Relevant source population}\label{fe_binary_pop}
Spectral lines formed in relativistic accretion discs are distorted due to Doppler and
relativistic effects producing a characteristic shape, which may be modelled to
determine the properties of the space time around the compact object
\citep{fabian89,laor91}. In particular, observations of supermassive black holes at
the center of galaxies have revealed broad skewed lines, which have allowed
constraints to be placed on the spin of the black holes
\citep{tanaka95,nandra97,nandra07,miller07,delacalle10}. 
The current status of our knowledge of relativistically broadened iron lines from 
AGN is summarized in \cite{guainazzi11}. 

\begin{figure}
\centerline{\psfig{file=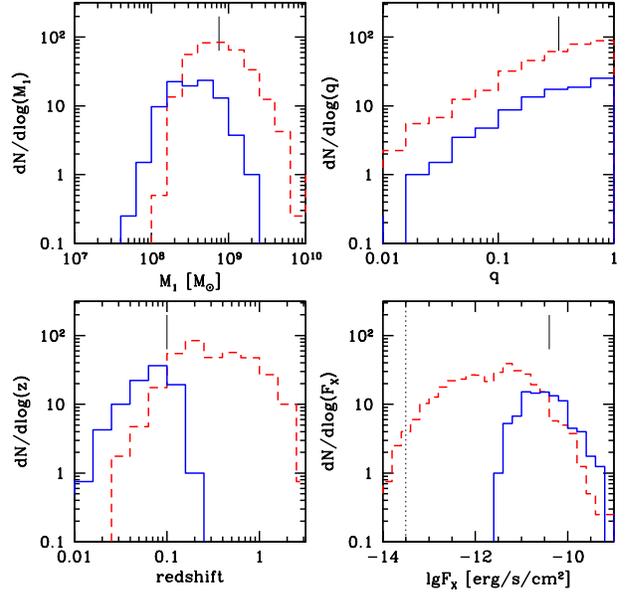,width=84.0mm}}
\caption{Properties of the individually resolvable PTA-MBHB population, assuming ten
  resolvable sources per frequency bin, and ten years of PTA observations, for our
  default model. All resolvable sources contributing at a level of 1ns or more to the
  PTA signal in the $3\times10^{-9}-10^{-6}$Hz frequency window are considered.  From
  the top left to the bottom right, we plot the primary mass $M_1$, mass ratio $q$,
  redshift and X-ray flux distributions. The vertical lines indicate the properties of
  the system studied in \S\ref{fe_lin_sim}. The total number of sources integrated
  over the blue histograms is $\sim 20$. Linestyle as in figure \ref{pop}.}
\label{resolvable}
\end{figure}

Double relativistic Fe K$\alpha$ lines are more likely to be
present in a 'steady' environment, where material rising from the accretion discs
(driven by heating and vertical stratification of the accreting material, magnetic
turbulence, winds, or star disc interaction) has time to shape a tenuous hot electron
plasma corona. They are therefore likely to appear in situations where $t_{\rm
  acc}>P_0$, but we do not exclude such possibility otherwise. Broad Fe K$\alpha$
lines appear to be common in AGN \citep{nandra07,delacalle10}, yet their
identification requires a large number of collected X-ray photons, i.e. deep,
targeted, time consuming observations. It is therefore reasonable to consider only
sources that may be individually resolvable in PTA campaigns, and consequently
localized in the sky to some accuracy \citep{SesanaVecchio10,corbin2010}. 

We therefore estimate the population of sources suitable for double Fe K$\alpha$ line
detection by considering individually resolvable sources only.  Unfortunately, the
concept of resolvability has not been deeply investigated in the PTA observation
context (and certainly not for eccentric binaries). In the circular binary case, a
rough 'one bin' rule estimate, provides $\sim10$ bright resolvable binaries
\citep{SVV09}. However, such estimate does not take into account the spatial
information enclosed in the detection with an array of pulsars; two sources at
different sky locations contribute differently in each pulsar, and their signals may
be disentangled even if their frequencies fall in the same bin. \cite{Boyle10}
estimated that, exploiting the spatial information enclosed in the signal, $2N/7$
sources per frequency bin would be resolvable by an array of $N$ pulsars. For our
estimate we therefore (somewhat arbitrarily) pick the ten strongest GW sources per
frequency bin, and impose a further cut at 1ns (dimmer sources would not be
individually detectable anyway). This leaves us with $\sim 100$ sources, the precise
number being vastly independent on the details of the global population model, but
only on the PTA observation time (assumed to be ten years).  Such population is shown
in figure \ref{resolvable} for our default model. Blue histograms represent MBHBs
still attached to their circumbinary discs (i.e., with $a>a^{\rm fr}$), and therefore
plausibly hosting small accretion minidiscs. We are left with $\sim 20$ bright, low
redshift sources.

Notice however, that our estimation of $a^{\rm fr}$ is quite conservative. In deriving
it, we equated the GW shrinking time to the viscous time of an {\it unperturbed} disc
at a radius corresponding to the inner rim of the gap. \cite{tanaka2010}, however,
showed that the steep density gradient at the inner edge of the disc will shorten the
inward diffusion timescale of the gas, causing a significant delay in the disc-binary
detachment, therefore increasing the number of sources with observable
minidiscs. Moreover, in the $\beta$-disc picture, the consumption time of a minidisc
of $\sim30R_S$ can easily exceed $10^4$ years (see equation (\ref{taccbeta})),
implying that a relevant fraction of the detached systems may still be significantly
accreting \citep{chang2010}.  It is in any case worth noticing that the relevant
population is sizable (maybe few-to-hundred sources) and probably mostly composed by
very low redshift systems ($z<0.2$, see lower left panel of figure \ref{resolvable}).

\subsection{Simulations of double K$\alpha$ line observability}\label{fe_lin_sim}
We assume the iron line to be similar to that observed from the archetype MCG-6-30-15
\citep{tanaka95,brenneman06,minutti07}.  We aim to assess the feasibility of detecting
pairs of relativistic iron lines which may be emitted from the inner minidiscs around
the merging black holes discussed herein, and to use them to constrain the properties
of the binary system, i.e., black holes spin, radial velocity, inclination etc.  In
order to investigate the feasibility of utilizing broad Fe K$\alpha$ emission from a
binary black hole merger, spectra were simulated with \textsc{xspec
  v12.6}\footnote{\url{http://heasarc.nasa.gov/xanadu/xspec/}} \citep{Arnaud96}. We
assume the availability of a next generation high throughput X-ray telescope, in
comparison to current instruments, i.e., \textit{XMM-Newton}. Specifically, we use the
response matrices created for the proposed \textit{Athena} mission
concept\footnote{\url{http://www.mpe.mpg.de/athena/home.php?lang=en}}.  The current
implementation envisions 2 focal plane instruments (i) the WFI a wide field CCD imager
with a 25 arcmin$^2$ field of view, and (ii) the XMS a narrow field of view (2.4
arcmin$^2$) calorimeter providing high resolution spectra in the Fe K$\alpha$ region
of 5eV. Hereafter, we assume that the binary system has been identified, opening the
possibility to study it in high spectral resolution with the XMS. 

\begin{figure}
\begin{center}
\includegraphics[height=0.35\textheight,angle=-90]{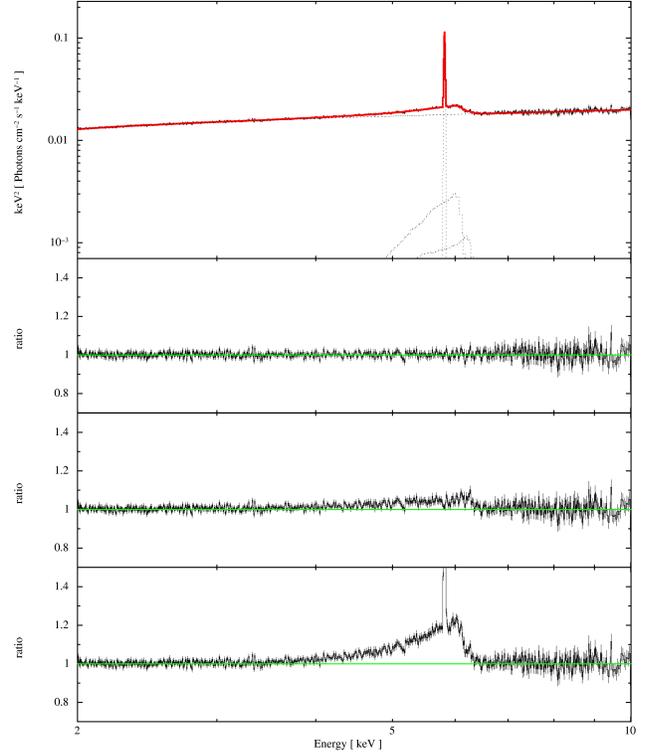}
\caption{Example of a blind fit model to the simulated spectrum described in
  \S\ref{fe_lin_sim}. The continuum is modeled with a power-law ($\Gamma \sim
  1.8$). Two relativistic lines in addition to a narrow line are required to provide
  an accurate fit. The second relativistic line is required at greater than the
  5$\sigma$ confidence level as measured by a simple F-test. The residual panels
  indicate no lines (bottom), a single broad line plus Gaussian (middle) and 2 broad
  lines plus Gaussian. See text for details.}
\label{fe_example_model}
\end{center}
\end{figure}

We assume a total mass of $\rm \sim 10^9 M_{\sun}$ for the binary pair and a mass
ratio of 1/3, consistent with the binary population presented in
Section \ref{fe_binary_pop}. A fiducial redshift of 0.1 and a bolometric luminosity of
10\% $\rm L_{Edd}$ and an associated bolometric correction of $\sim$ 20
\citep{vasudevan09,Lusso10}, i.e., $\rm f_{x1} \sim 10^{-11}~erg~s^{-1}~cm^{-2},
f_{x2} \sim 3 \times 10^{-11}~erg~s^{-1}~cm^{-2}$. The parameters of our fiducial
system are marked by vertical ticks in figure \ref{resolvable}.  The black hole binary
is assumed to have a periodicity consistent with a velocity separation of $\rm
10^4~km~s^{-1}$.  For simplicity, we assume the energy of the first iron line to be
consistent with 6.4 keV emission at a redshift of 0.1, while the second line is
blueshifted by $\rm 10^4~km~s^{-1}$ relative to the first line.

In order to simulate the expected X-ray spectrum from such a system, we first make a
number of simplifying assumptions. The broadband continuum emission from each
accreting black hole is assumed to have a power-law shape, where the spectral index
has a value of $\Gamma = 1.8$. The relativistic iron line is modelled using the
\texttt{laor} line profile \citep{laor91}, where the equivalent width of the line is
set to be approximately 150eV, as expected for reflection from a disc surrounding a
black hole \citep{george91}. The emissivity profile of the disc is fixed at $\rm
R^{-3}$, and the outer radius of the emitting region is assumed to be 40 $\rm R_S$ (see
\S\ref{subsec:lightcurves}).  A narrow line consistent with Fe K$\alpha$ emission from
material at much larger radii is also included. These values are consistent with
current available observations \citep[e.g.,][]{delacalle10,guainazzi11}.  A more
thorough and self-consistent modelling of the expected spectrum from a MBHB system,
for example, including complex absorbers and self consistently calculating the
continuum plus reflected emission is beyond the scope of this work (e.g., see
\citealt{brenneman11}), and as such we defer a detailed analysis of this problem for
future investigations.

\begin{figure*}
\begin{center}
\subfigure{\includegraphics[height=0.35\textwidth]{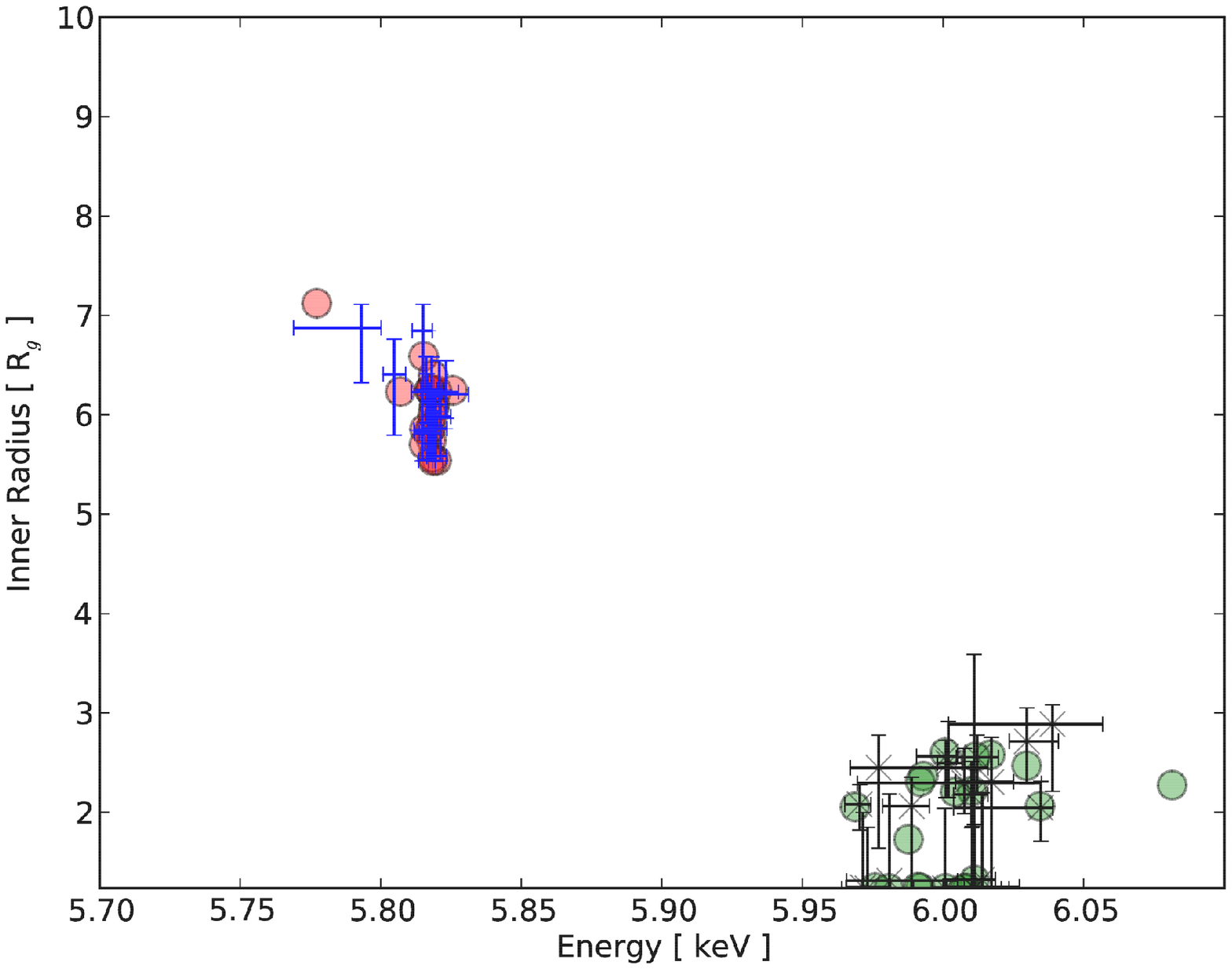}}
\subfigure{\includegraphics[height=0.35\textwidth]{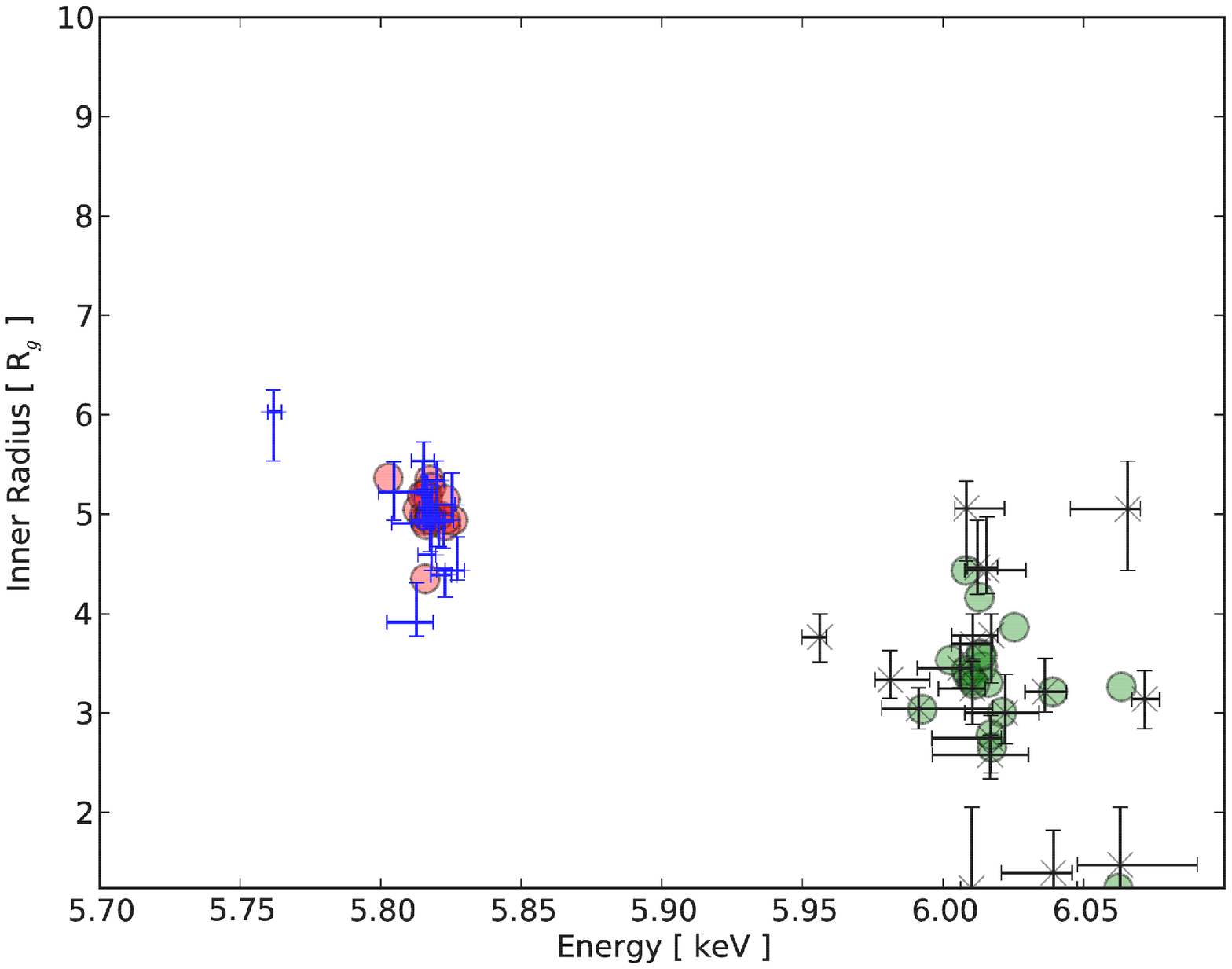}}
\caption{\textbf{Left:} Results of the best fit model in figure \ref{fe_example_model}
  for binary system with spins of 0.0 \& 0.9 at an inclination of
  30$\degr$ for a random sample of 20 model realisations. \textbf{Right:} As in the
  figure on the left but here the black hole spins are set to 0.3 \& 0.7
  respectively at an inclination of 30$\degr$. The colored points indicate the results
  of fits initialized at the best fit values, whereas the plus \& cross symbols
  indicate the results of the best fit blind model. Error bars (90\% confidence level)
  are plotted on the blind model results only for clarity. The exposure time is
  200ks. In both cases the lines are resolved.}
\label{fe_fit_results}
\end{center}
\end{figure*}

The model described above was defined in \textsc{xspec} as
\texttt{pha*((zpo+laor)+(zpo+laor)+ gauss)} and 200 spectra were simulated in each
case, using the latest available response matrices. An example of a simulated spectrum
is shown in the top panel of figure \ref{fe_example_model}. The interstellar
absorption component is modelled via the \texttt{pha} model; however, as we are
interested in the Fe line region, modest values for the column density (i.e., $\rm
\lesssim 10^{22}~cm^{-2}$) will have a negligible effect. Hence the column density was
held constant in all fits and as such is not discussed further. The spectral index
$\Gamma$ is assumed to be identical in both sources as it is inherently difficult to
accurately extract both indices correctly, if the difference between them is small,
due to the narrow assumed bandpass (0.5 -- 10.0 keV).  After each spectrum was
simulated the \texttt{fit \& error} commands were then run in order to measure the
best fit parameters given the signal-to-noise (SNR) of the spectrum. 
This results in scatter around the
defined model which is indicated in color in figure \ref{fe_fit_results}. This model was
then removed and a new 'blind' model was defined. In this case only 2 parameters are
known a priori, the redshift and the line of sight column density. Additionally the
inclination of the line is fixed at the known value.  All other parameters are
initialized at reasonable values given the observed spectrum. The model is first fit
with only a single relativistic line and the narrow Fe K$\alpha$ line
(\texttt{pha*(zpo+laor+zgauss)}). After the best fit is obtained, a second
relativistic line is added to the model and the significance of this line is
calculated using a simple \texttt{ftest}, e.g., see figure \ref{fe_example_model}. In
all cases the inclination of both lines are tied to each other, as we expect the
angular momentum of each black hole to be aligned by this time in the merging process
\citep{Bogdanovic:2007hp,Dotti:2009vz}.

We focus on 2 primary cases for the spin of the merging black holes (i) $\rm a_1 =
0.0,~a_2 = 0.9$, and (ii) $\rm a_1 = 0.3,~a_2 = 0.7$.  In figure \ref{fe_example_model},
we plot an example best fit model to one of our simulated spectra in case (i)
above. Here the inclination is 30$\degr$.  In figure \ref{fe_fit_results}, we display
the results of the fit to the relativistic iron lines. The input model is in colour,
whereas the subsequent blind fits are indicated by the plus and cross symbols. The
exposure time in this case is 200ks. It is clear that it will be possible to resolve
and constrain both Fe lines to high accuracy. Similar results are achieved for case
(ii). At higher inclinations our ability to accurately recover the line parameters
deteriorates as expected due to the relative narrowing of the line profile. We note
that in all of the lower inclination models ($\lesssim 40\degr$), all of the blind
fits require the presence of two relativistic lines at greater than the 5$\sigma$
confidence level. As we move to higher inclinations, this decreases somewhat whereby
at $60\degr$ only $\sim$ 95\% of our realisations require 2 relativistic lines.

The primary uncertainty in all models is the inclination of the binary system, due to
the degeneracy in the line shape with changing inclination and/or spin. 
If we relax the initial inclination constraint imposed above, our ability to
accurately determine the line parameters is weakened. For example in case (i) above,
a second line is required at $\gtrsim 5\sigma$ level on only $\sim$ 85\% of our
models, while only 95\% require a second line at the 3$\sigma$ level.
We expect, however, to get some indicative prior on the system inclination from
PTA observations. Infact, \cite{SesanaVecchio10} showed that the GW signal analysis 
will enable a determination of the system inclination within a $\sim$20 deg accuracy
(assuming a source SNR of 10). 
Models were also created with additional narrow emission/absorption lines consistent
with Fe XXV/XXVI. As in the case of the narrow 6.4 keV line, the exquisite resolution
provided by the calorimeter in the Fe K region allows these lines to be easily
resolved. We also experimented with decreasing the velocity separation of the broad
line components; however, in this case the results are strongly inclination
dependent. The use of a more accurate relativistic line profile model, e.g.,
\texttt{kerrdisk} \citep{brenneman06}, would help in this case.

There are a number of caveats which should be considered when interpreting the results
above. For example, even in the case of MCG-6-30-15 where long high SNR observations
exist, there is uncertainty in the parameters of the observed line when modelled by
different groups \citep{fabian02,reynolds04,brenneman06}. In addition, complex
relativistic effects (e.g., light bending, \citealt{miniutti04}) may render the line
indistinguishable from the continuum in the absence of highly sophisticated spectral
modelling \citep[e.g.,][]{bhayani11}.  Finally, we note that an alternative physical
model has been proposed to explain the relativistic lines observed in numerous AGN. In
this scenario, the lines are produced by a combination of non-relativistic fluorescent
line emission from distant material and complex absorption by material in the inner
accretion flow, e.g., for further details see the review by \citet{turner09}.

\section{Discussion and conclusions}
The distinctive signatures of PTA-MBHBs highlighted above open interesting scenarios
for combining pulsar timing and X-ray observations in the coming years. Even though
current PTA efforts (PPTA, EPTA, NANOGrav) may eventually succeed in the detection
challenge, multimessenger astronomy prospects are particularly promising in the
context of the planned SKA. If a nominal 1ns sensitivity level will be achieved, SKA
will make GW astronomy with pulsar timing possible. Several hundreds of signals
emitted by the low redshift population of MBHBs will add up to form a confusion noise,
on top of which some (possibly a hundred) sources will be individually resolvable and
their position in the sky determined.

In this context, X-ray (but also other bands, not considered in this paper)
observations may play either a preparatory or a follow-up role. On one hand, the
detection of a population of periodic X-ray sources in the present decade, may provide
useful information for PTA observations. By the time the SKA will be online, we may
have identified a catalogue of systems in the sky from which we expect to detect GW
signals, being also able to give indications about the expected frequency. This may
substantially facilitate the PTA detection and analysis pipelines, by reducing the
search parameter space for specific signals. On the other hand, several resolvable
sources with high SNR may be located in the sky with enough accuracy to make
follow-up X-ray monitoring possible. To be conservative, according to
\cite{SesanaVecchio10}, typical PTA source sky location accuracy is expected to be in
the order of tens of square degrees for a source SNR$=10$ \citep[but, under specific
  detection conditions, it can actually be an order of magnitude better,
  see][]{corbin2010}. According to figure \ref{resolvable}, typical masses are $>{\rm
  few}\times10^{8}\msun$ at $z<0.3$. Such systems should be host in massive galaxies
with stellar bulges of masses $>10^{11}\msun$ \citep{gultekin09}, whose number
density is $\lesssim 10^{-3}$Mpc$^{-3}$ \citep{bell2003} Considering a sky location
error of 10 deg$^2$, the comoving volume enclosed in the error box is $2\times
10^{-3}$Gpc$^{3}$: maybe up to few thousands candidates falls in the error-box.  Since
the presence of a distinctive counterpart relies on the assumption of accretion,
active galaxies only should be selected, which may leave us with few dozens of
candidates.  Further down selection may be performed by keeping galaxies that show
signatures of recent merger activity \footnote{A detailed study of the statistics
of candidate hosts can be found in the independent study by \cite{tanaka11}, which
is complementary to ours in several aspects.}. If just one or at most a few systems are
identified, ultra deep X-ray exposure may be used to reveal characteristic double
relativistic Fe K$\alpha$ emission lines. 

Such investigation will likely be possible in the next decade with future generation
of pulsar timing arrays (in particular with the SKA) and X-ray observatories. In this
paper we quantified for the first time the population of expected sources, and their
characteristic signatures, and we proposed strategy for coordinating multimessenger
observations. We summarize in the following our main results:
  
\begin{itemize}
\item About a thousand MBHBs are expected to contribute to the PTA detectable signal 
in the frequency range $10^{-9}-10^{-6}$Hz, at a level of 1ns or more. Uncertainties 
in the MBHB population model and in the detailed dynamics of the contributing systems
may impact such figures by a factor of a few. 

\item We assumed the standard picture of MBHB migration in circumbinary discs. For
  popular geometrically thin, optically thick discs described by typical parameters,
  we find that many of these PTA sources (order of few hundreds, 20\%-to-50\% of the
  total population, for $0.1<\alpha<0.3$ and $0.1<\dot{m}<0.3$) are coupled to their
  circumbinary discs, making a strong case for looking to possible electromagnetic
  signatures rising by the disc-binary dynamical interplay.

\item Detailed SPH simulations of the eccentric binary-disc interaction highlight
  periodic streams of gas leaking from the inner edge of the circumbinary disc,
  through the low density central gap. The rate at which such streams feed the central
  binary can be a significant fraction of the Eddington accretion rate of the MBHB
  (after initial relaxation of the system, we find an average
  $\dot{m}\approx0.4$). The streaming periodicity is extremely sharp, with a variation
  in the MBHB feeding rate of a factor of two-to-four.

\item By modelling analytically the dynamics and the emitted spectrum of the
  inflow-fed minidiscs forming around the two MBHs, we identified several
  observational features that may be distinctive of an accreting MBHB. We concentrated
  on the X-ray domain, by separately studying X-ray periodicity and double
  relativistic Fe K$\alpha$ lines.

\item Assuming all merging MBHBs are accreting, we estimate about 100-500 (depending
  on the detailed property of the circumbinary disc) periodically variable X-ray
  sources with periods between one and five years. The observed flux on Earth for most
  of the sources is larger than $10^{-13}$erg s$^{-1}$cm$^{-2}$, within the sensitivity reach
  of the upcoming X-ray all sky monitor eRosita. However, a lightcurve with at least
  15-20 points (eRosita will point each region of the sky at 8 different times only)
  is required for a statistically significant detection of such a periodicity.

\item Iron line identification requires deep, targeted observations. We therefore
  consider as suitable candidate only those PTA sources that are individually
  resolvable, for which the sky location can be identified to some accuracy. The
  number of suitable targets is of the order of few tens, depending both on the
  ability of PTA of resolving sources in the sky, and on the details of the disc-MBHB
  decoupling and the physical nature of the minidiscs.

\item Assuming that the accretion flow onto each black hole is capable of generating
  relativistic Fe K$\alpha$ emission lines, we have demonstrated that it will in
  principle be feasible, via high spectral resolution observations with a next
  generation X-ray observatory (e.g. {\it Athena}), to identify double K$\alpha$ line
  features. Such features can be used to estimate the properties of the black holes,
  forecasting, in combination with modelling of the PTA signal, the possibility to
  constrain the space-time around the black hole to unprecedented accuracy.

\item Note that we assumed {\it all} MBHB to be surrounded by a circumbinary disc in
  their late evolutionary stage, prior to merger. Even though this is likely for gas
  reach merging systems, it might be a too extreme assumption for the massive, low
  redshift sources relevant to our study. Even assuming that only 10\% of the systems
  are surrounded by a circumbinary disc, the number of detectable sources is still
  interesting: about 10-50 bright, periodic X-ray sources; at least a few individually
  resolvable PTA sources showing double Fe K$\alpha$ line profiles.

Even though GW detection of MBHBs remains a challenging task for the present and
future astrophysical generations, systematic pulsar timing campaigns are ongoing, and
their accuracy and sensitivity will inevitably improve in the coming years. Pulsar
timing will therefore provide a safe, open GW window on the low frequency
Universe. The combination with electromagnetic observations, such the ones proposed in
this paper, will help exploiting GW detection capabilities at their best, providing a
lot of information about the population and dynamics of MBHBs.  The present paper is
just a first step into the realm of multimessenger astronomy with pulsar timing, hoping
that investigators from both the GW and the X-ray/optical/radio communities will take
the challenge following our footsteps.

\end{itemize}
\section{Acknowledgments}
As we were completing this work, we became aware of a concurrent
independent study by \cite{tanaka11} 
addressing similar questions.
We thank J. Krolik and Margherita Giustini for insightful discussions,
and Francesco Haardt and Elena Rossi for the detailed comments on the manuscript.
CR wishes to thank Jorge Cuadra for his modified version of Gadget-2.

\bibliographystyle{mn2e}
\bibliography{aeireferences,fe_line}

\end{document}